\documentclass[prd,
              preprint,
              superscriptaddress,
              preprintnumbers,
              eqsecnum,
              showpacs,
              nolongbibliography,
              nofootinbib,
              nobibnotes,
              noeprint
              ]{revtex4-2}

\usepackage[linkcolor=blue,
            citecolor=blue,
            urlcolor=blue,
            colorlinks=true,
            breaklinks
            ]{hyperref}

\usepackage{float}
\usepackage{booktabs}
\usepackage[italic]{hepnames}
\usepackage{amsfonts}
\usepackage{amsthm}
\usepackage{bm}
\usepackage{slashed}
\usepackage{dsfont}
\usepackage{tikz}
\usepackage[shortlabels]{enumitem}
\usepackage{xspace}
\usepackage{siunitx}
\usepackage{hyperref}
\usepackage{bookmark}
\usepackage{bbm}

\makeatletter
\newcommand{\defineshort}[2]{%
  \expandafter\newcommand\csname #1\endcsname{%
    #2%
    \futurelet\@tempnext\@acr@spacecheck
  }%
}
\def\@acr@spacecheck{%
  \ifx\@tempnext.\else
    \ifx\@tempnext,\else
      \ifx\@tempnext;\else
        \ifx\@tempnext:\else
          \ifx\@tempnext!\else
            \ifx\@tempnext?\else
              \space
            \fi
          \fi
        \fi
      \fi
    \fi
  \fi
}
\makeatother

\defineshort{wtig}{$[\rm WTI]_{\Gamma_{\!5}}$}
\defineshort{wtiG}{$[\rm WTI]_{G_5}$}
\defineshort{opr}{1PR}
\defineshort{opi}{1PI}

\def\1eq#1{Eq.\nobreak\thinspace(\ref{#1})}

\def\2eqs#1#2{Eqs.\nobreak\thinspace(\ref{#1}) and\nobreak\thinspace(\ref{#2})}
\def\3eqs#1#2#3{Eqs.\nobreak\thinspace(\ref{#1}),\nobreak\thinspace(\ref{#2}) and\nobreak\thinspace(\ref{#3})}

\def\fig#1{\hyperref[#1]{Fig.\nobreak\thinspace\ref*{#1}}}

\def\sect#1{\hyperref[#1]{Sec.\nobreak\thinspace\ref*{#1}}}
\def\appref#1{\hyperref[#1]{App.\nobreak\thinspace\ref*{#1}}}

\def\ie{{\it i.e.}, }
\def\eg{{\it e.g.}, }

\newcommand{\be}{\begin{equation}}
\newcommand{\ee}{\end{equation}}

\newcommand{\bea}{\begin{eqnarray}}
\newcommand{\eea}{\end{eqnarray}}

\newcommand{\sla}{\slash \hspace{-0.22cm}}

\def\s#1{{\scriptscriptstyle #1}}

\def\is{S^{-1}}             % Inverse quark propagator
\def\g{\Gamma}              % Short hand for \Gamm
\def\go{\Gamma_{\!1}}       % Vertex part with odd # of Dirac structures
\def\ge{\Gamma_{\!2}}       % Vertex part with even # of Dirac structures

\def\ga{\Gamma_{\!5}}

\def\GNEW{G_{\!5\mu\nu}^{({\bf \s 1})}}

\begin{document}

%%%%%%%%%%%%%%%%%%%%%%%%%%%%%%%%%%%%%%%%%%%%%%%%%%%%
% Title & Author list
%%%%%%%%%%%%%%%%%%%%%%%%%%%%%%%%%%%%%%%%%%%%%%%%%%%%
\title{\boldmath A comprehensive approach to the physics of mesons}

\author{A.~S.~Miramontes}
\email{angel.s.miramontes@uv.es}
\affiliation{\mbox{Department of Theoretical Physics and IFIC, University of Valencia and CSIC}, E-46100, Valencia, Spain}

\author{J.~M.~Morgado}
\email{jose.m.morgado@uv.es}
\affiliation{\mbox{Department of Theoretical Physics and IFIC, University of Valencia and CSIC}, E-46100, Valencia, Spain}

\author{J.~Papavassiliou}
\email{joannis.papavassiliou@uv.es}
\affiliation{\mbox{Department of Theoretical Physics and IFIC, University of Valencia and CSIC}, E-46100, Valencia, Spain}
\affiliation{\mbox{ExtreMe Matter Institute EMMI, GSI, Planckstrasse 1, Darmstadt, 64291, Germany}}

\author{J.~M.~Pawlowski}
\email{j.pawlowski@thphys.uni-heidelberg.de}
\affiliation{\mbox{Institut f\"ur Theoretische Physik, Universit\"at Heidelberg}, Philosophenweg 16, Heidelberg, 69120, Germany}
\affiliation{\mbox{ExtreMe Matter Institute EMMI, GSI, Planckstrasse 1, Darmstadt, 64291, Germany}}

%%%%%%%%%%%%%%%%%%%%%%%%%%%%%%%%%%%%%%%%%%%%%%%%%%%%
% Abstract
%%%%%%%%%%%%%%%%%%%%%%%%%%%%%%%%%%%%%%%%%%%%%%%%%%%%
\begin{abstract}

We develop a novel approach for the self-consistent solution of coupled sets of Bethe-Salpeter and Schwinger-Dyson equations in QCD. This framework allows us to maintain the axial  
Ward-Takahashi identities 
of the theory within advanced approximation schemes, such as the skeleton or three-particle irreducible expansions. 
For this purpose we reformulate   
the Schwinger-Dyson equation of the 
axial-vector vertex such  
that the bulk of its quantum corrections 
is expressed 
in terms of a novel vertex. Crucially, this vertex 
satisfies a symmetry-induced relation of its own, which involves the full quark-gluon vertex.
As a result, the  Schwinger-Dyson equation reproduces 
the standard Ward-Takahashi identity
satisfied by the axial-vector vertex. Consequently, 
the known relation
between the quark mass function
and the wave function of the pion in the chiral limit is duly fulfilled. 
The present approach offers valuable insights into the interplay between 
symmetry and dynamics, and provides a practical path towards computations of hadron physics within sophisticated approximations. 
In particular, the one-loop dressed truncation of the key dynamical equations, including that of the quark-gluon vertex, is shown to be completely compatible with  the required symmetry relations. Further extensions and 
potential phenomenological applications of the developed 
framework are briefly discussed.

\end{abstract}

\maketitle

\newpage 

%%%%%%%%%%%%%%%%%%%%%%%%%%%%%%%%%%%%%%%%%%%%%%%%%%%%
% Main body
%%%%%%%%%%%%%%%%%%%%%%%%%%%%%%%%%%%%%%%%%%%%%%%%%%%%
\section{Introduction}\label{sec:Intro}

In recent years, our quantitative understanding of the correlation functions of QCD 
has advanced considerably, mainly due to the 
ongoing efforts of functional approaches, such as 
Schwinger-Dyson equations (SDEs), for reviews see  \eg\cite{Roberts:1994dr, Alkofer:2000wg, Maris:2003vk, Fischer:2006ub, Roberts:2007ji, Binosi:2009qm, Maas:2011se, Cloet:2013jya, Eichmann:2016yit, Fischer:2018sdj, Huber:2018ned, Ferreira:2023fva}, and the 
functional renormalization group (fRG), 
for QCD-related reviews see \eg\cite{Litim:1998nf, Berges:2000ew, Pawlowski:2005xe, Schaefer:2006sr, Gies:2006wv, Braun:2011pp, Dupuis:2020fhh, Fu:2022gou}. In fact, by now, the results for quark-gluon correlation functions from functional approaches are in excellent agreement with that from gauge-fixed lattice simulations. 

However, the incorporation of 
this extensive knowledge into the physics of hadrons is far from straightforward, chiefly due to incompatibilities  
between standard truncation schemes and the underlying fundamental symmetries,
expressed through the Ward-Takahashi identities (WTIs).  
In fact,  
to date, 
most applications of functional approaches to hadronic physics are still based on the rainbow ladder (RL) approximation \cite{Maris:1999nt, Maris:1999bh, Alkofer:2002bp, Eichmann:2008ae, Qin:2011dd, Hilger:2014nma, Hilger:2015hka, El-Bennich:2016qmb, Mojica:2017tvh, Raya:2017ggu, Weil:2017knt, Serna:2017nlr, Hernandez-Pinto:2023yin, Hernandez-Pinto:2024kwg}; for works beyond RL, see, \eg  \cite{Munczek:1994zz, Matevosyan:2006bk, Fischer:2007ze, Fischer:2008wy, Williams:2014iea, Sanchis-Alepuz:2014wea, Williams:2015cvx, Sanchis-Alepuz:2015qra, Binosi:2016rxz, Williams:2018adr, Miramontes:2021xgn, Miramontes:2022mex, Gao:2024gdj, Miramontes:2025ofw, Fu:2025hcm}.  
This method and its variants \cite{Chen:2019otg,Chang:2020iut,Xu:2024fun, Xu:2025hjf,Xu:2024vkn}
encompass a special type 
of QCD-derived information, 
typically amassed into     
propagator-like constructs 
(\eg effective charges~\cite{Aguilar:2009nf,Binosi:2014aea,Cui:2019dwv,Deur:2023dzc,Gao:2024gdj}), but do not include the information  
encoded in the dressings of the fundamental QCD vertices, and in particular, of the quark-gluon vertex,  
$\Gamma_{\mu}$. 

In general, the physics of hadrons involves a large set of dynamical equations, non-trivially coupled to each other \cite{Roberts:1994dr,Maris:2003vk,Cloet:2013jya}. 
In particular, the standard SDEs for the quark-gluon correlation functions must be combined with the appropriate bound-state equations,  
namely the Bethe-Salpeter equations (BSEs) in the case of mesons \cite{Maris:1997tm,Sanchis-Alepuz:2015tha}, or the Faddeev equations in the case of baryons \cite{Nicmorus:2008vb,Eichmann:2009qa,Eichmann:2011vu,Heupel:2012ua,Sanchis-Alepuz:2011egq,Sanchis-Alepuz:2013iia,Wallbott:2019dng,Yao:2024ixu,Eichmann:2025etg,Eichmann:2025gyz}.  
Focusing on mesons, 
in order for a self-consistent 
solution to emerge, any approximation implemented at the 
level of the SDEs must be appropriately 
incorporated into the BSEs. 
In fact, a pivotal consistency requirement 
of any truncation scheme 
is that the WTIs \cite{Ward:1950xp,Takahashi:1957xn, Fujikawa:1980eg,Itzykson:1980rh,Miransky:1994vk}
be exactly preserved, in order for the chiral dynamics to be faithfully captured.  
The tension within the 
existing approaches may be summarized by stating that 
the proper treatment of the 
quark gap 
equation 
is known to 
require quite elaborate 
ingredients, whose incorporation into the BSEs destabilizes the WTIs. Therefore, it is highly desirable to set up functional approaches that allow for self-consistent computations, by including or computing state-of-the-art QCD correlation functions. This would allow us to use the respective quantitative functional results from \eg\cite{Gies:2002hq, Pawlowski:2003hq, Fischer:2004uk, LlanesEstrada:2004jz, Alkofer:2008tt, Windisch:2012de, Hopfer:2012cnq, Mitter:2014wpa, Williams:2014iea, Aguilar:2014lha, Braun:2014ata, Rennecke:2015eba, Aguilar:2016lbe, Cyrol:2017ewj, Aguilar:2018epe, Oliveira:2018fkj, Oliveira:2018ukh, Albino:2018ncl, Fu:2019hdw, Tang:2019zbk, Huber:2020keu, Oliveira:2020yac, Gao:2021wun, Albino:2021rvj, Aguilar:2024ciu, Fu:2025hcm}, as well as 
results from gauge-fixed lattice simulations, see \eg \cite{Skullerud:2002ge,Skullerud:2003qu, Skullerud:2004gp, Ilgenfritz:2006he, Kizilersu:2006et, Bogolubsky:2007ud, Bogolubsky:2009dc, Oliveira:2009eh, Oliveira:2010xc, Ayala:2012pb, Sternbeck:2012mf, Bicudo:2015rma, Duarte:2016iko, Sternbeck:2017ntv, Boucaud:2017obn, Aguilar:2019uob, Aguilar:2021lke, Kizilersu:2021jen, Skullerud:2021pel, Chang:2021vvx, Pinto-Gomez:2022brg, Pinto-Gomez:2024mrk}. 

Given the key r\^ole played by the WTIs, in the present work we shall use chiral self-consistency as our main guiding principle. In particular, the novel element we introduce 
is to enforce {\it exactly}
the WTI satisfied by the flavour non-singlet axial-vector vertex, $\ga^{\mu}$,
at the level of the SDE that 
governs this vertex, employing the form first presented in \cite{Bender:2002as,Bhagwat:2004hn}. This is 
carried out 
first at the level of 
the complete SDE, and then within its 
``one-loop dressed'' approximation.
It turns out that the success 
of this endeavor hinges on 
the use of a 
quark-gluon vertex that contains 
nontrivial dressings  for all of its form factors.
In fact, quite importantly,  
these dressings cannot be arbitrary;
instead, they must  
be determined {\it dynamically},  
as solutions of a standard version of 
the SDE that controls the evolution of $\Gamma_{\mu}$.

\begin{figure}[!t]
    \centering
    \includegraphics[scale=1]{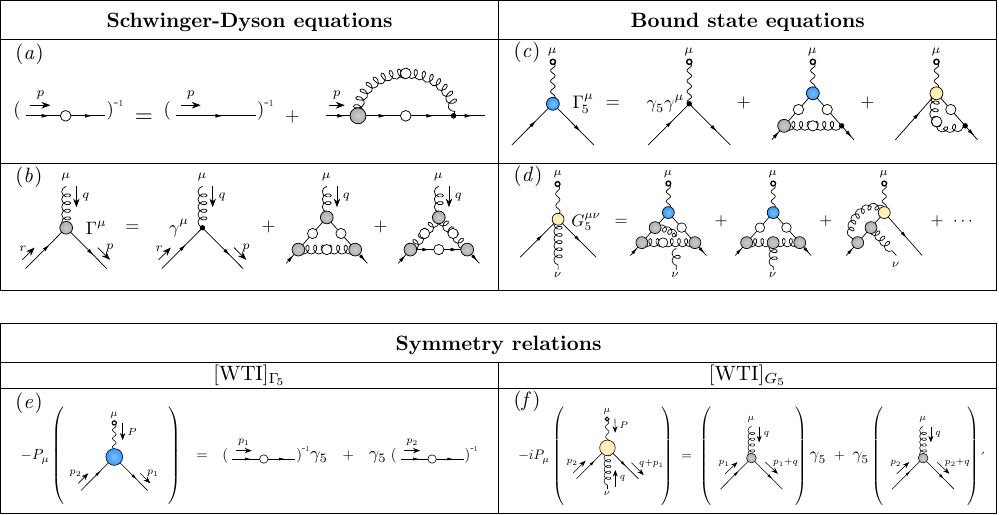}
    \caption{Summary of the main building blocks of this work, shown 
    graphically in panels (${\it a}$)-(${\it f}$):\\ 
    (${\it a}$) quark gap equation; (${\it b}$)  one-loop dressed version of the quark-gluon vertex SDE; (${\it c}$) axial-vector vertex SDE; (${\it d}$) one-loop dressed version of the gluon-axial-vector vertex SDE; (${\it e}$) axial-vector vertex WTI; and (${\it f}$) gluon-axial-vector vertex WTI.}
    \label{fig:summary}
\end{figure}

\begin{figure}[t!]
    \centering
    \includegraphics[scale=1]{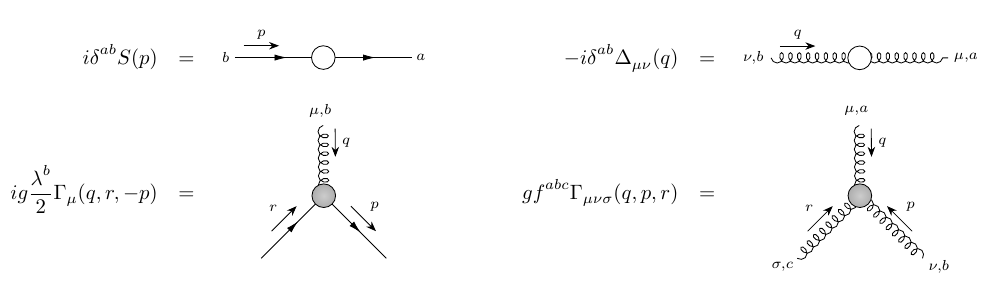}
    \caption{Diagrammatic representation of the elementary Green functions appearing in this work: \textit{top-left}: quark propagator; \textit{top-right}: gluon propagator; \textit{bottom-left}: quark-gluon vertex; and \textit{bottom-right}: three-gluon vertex.}
    \label{fig:Elementary}
\end{figure}

In the remainder of this section we 
briefly elucidate the sequence of ideas 
followed in this work, and highlight 
the main results.
To that end, we employ a  
visual overview of the key 
components of this analysis, shown in the 
panels ({$\it a$})-({$\it f$})
of \fig{fig:summary}. In particular, we have 
\begin{enumerate}[topsep=0pt, itemsep=1pt, parsep=0pt, partopsep=0pt,label=(\textit{\alph*})]
        \item 
        the quark propagator, $S(p)$,
        and the SDE 
        (gap equation) that determines its 
        momentum dependence;
        a key ingredient 
        of this equation is the
        fully-dressed quark-gluon vertex, $\Gamma_{\!\mu}$, discussed in the  
        next panel. 
        \item the quark-gluon vertex, $\Gamma_{\!\mu}(q,r,-p)$, and the one-loop dressed version of the corresponding SDE, obtained within the formalism of the three-particle-irreducible (3PI) effective action, at the three-loop order~\cite{Alkofer:2008tt,Williams:2015cvx,Aguilar:2024ciu}. 
        \item the standard non-singlet axial-vector vertex, $\ga^{\mu}(P, p_2,-p_1)$, 
        and the exact closed form of the SDE that governs it.  In the limit $P \to 0$, this SDE collapses to the BSE that describes  the properties of the  pion.   
        The second diagram on the r.h.s. is the standard RL contribution (with $\Gamma_{\mu} \to \gamma_{\mu}$). The third diagram is composed by 
        a vertex denoted by  
        $G_5^{\mu\nu}$, first
        introduced in \cite{Chang:2009zb} under the name $\Lambda_5^{\mu\nu}$,
        which captures {\it all} remaining contributions (see next panel);
        we refer to it as the {\it ``gluon-axial-vector vertex''}. 
        Note that $G_5^{\mu\nu}$ contains 
        a pole in $P^2$, whose residue is 
    related to the quark-gluon vertex. 
        \item 
       the dynamical equation that determines the 
        gluon-axial-vector vertex, $G_5^{\mu\nu}$, in the one-loop dressed 
        approximation.   
        \item 
        the axial WTI satisfied by 
        $\ga^\mu$, denoted for 
        brevity by \wtig, 
         which involves a special combination of the full inverse quark propagator.
        \item 
        the axial WTI satisfied by the vertex 
        $G_5^{\mu\nu}$, called $[{\rm WTI}]_{G_5}$;  
        it is akin to that of $\ga^\mu$, but involves the full quark-gluon vertex, $\g_{\mu}$, instead of the inverse quark propagator. We emphasize that this WTI corresponds to relation stated in  Eq.~6 of  \cite{Chang:2009zb}.
\end{enumerate}
For the benefit of the reader, 
in \fig{fig:Elementary} we summarize the 
correlation functions entering in our analysis. Note also that throughout this 
article we use conventions and Feynman rules written in Minkowski space, see \eg\cite{Itzykson:1980rh,Binosi:2009qm}.

The first nontrivial result 
presented in this work 
is the 
demonstration  that the \wtig holds true 
at the level of the {\it full} SDE 
that governs the vertex 
$\ga^{\mu}$, see panel  (${\it c}$). Note that, 
crucially, the \wtig involves the quark propagator, whose 
self-energy contains the full 
$\Gamma_{\mu}$, see panel  (${\it a}$). 
This demonstration  becomes 
possible precisely because of the inclusion of the  vertex $G_5^{\mu\nu}$, 
contained in the 
third graph on the r.h.s. of 
panel  (${\it c}$); in particular, 
$G_5^{\mu\nu}$
eliminates  exactly a symmetry violating contribution stemming from the second graph. We emphasize that the proof hinges on the validity of \wtiG, shown in panel (${\it f}$).

There is an important 
consequence of this result, 
which brings us one step closer to the 
BSE satisfied by the pion. In particular, 
it is well-known that the dynamical breaking of the chiral symmetry is 
encoded in the emergence of a nonvanishing mass function, $B(p^2)$, 
which appears  on the r.h.s. of the 
\wtig, panel (${\it e}$). The only way to reconcile this  feature 
 is by allowing for 
 a massless pole in $\ga^{\mu}$, whose 
 residue is essentially the pion wave function. In the limit $P \to 0$, 
 the main component of the pion 
 wave function satisfies a fundamental 
 relation, connecting it directly 
 to the quark component 
$B(p^2)$. As we show in this work, 
the BSE satisfied by this  
component, derived as the 
$P \to 0$ of the SDE governing  
$\ga^{\mu}$, panel (${\it c}$),
reduces precisely to the dynamical equation obtained for
$B(p^2)$ from the gap equation in (${\it a}$),  
with $\Gamma_{\mu}$
fully-dressed.
We stress that, at this point, $\Gamma_{\mu}$ is complete; in particular, it is 
{\it not} approximated by the SDE 
in panel (${\it b}$), which has not yet been employed.  

Since, for practical purposes, 
the SDE of $\ga^{\mu}$
may not be treated 
exactly, it is of the utmost importance to implement a truncation 
that would respect an 
approximate version of 
the aforementioned key 
properties. 
This brings us to one of the major highlights of this work.
In particular,   
the one-loop dressed version of $G_5^{\mu\nu}$ in panel (${\it d}$) 
satisfies a very concrete 
version of the 
\wtiG in panel (${\it f}$):
the $\g_\mu$ that emerges 
on the r.h.s. of the \wtiG
{\it must} solve the 
one-loop dressed 
version of its own SDE, shown in 
panel (${\it b}$)! 
Thus, the dynamics and symmetries captured by all panels of 
\fig{fig:summary}
are mutually compatible and 
coherently intertwined. 

The article is organized as follows. In \sect{sec:GapEq}, we review 
certain key aspects of the quark propagator 
and the quark-gluon vertex, and discuss the form of the 
SDEs that govern their evolution. 
Next, in \sect{sec:WTI}, we present the 
WTIs of the two  
axial-vector vertices, $\ga^{\mu}$ and 
$G_5^{\mu\nu}$, and their  respective pole contents,  imposed by the dynamical breaking of the chiral symmetry. 
In \sect{sec:SDE} 
we demonstrate that
the SDE of the $\ga^{\mu}$ 
reproduces the correct WTI, 
provided that the 
vertex $G_5^{\mu\nu}$ is properly included in the SDE kernel. 
In \sect{sec:ChiralLim}, we show that, in the chiral limit, 
the previous SDE 
reproduces exactly the fundamental 
relation between the 
pion Bethe-Salpeter Amplitude (BSA) and the quark mass function.
In \sect{sec:OpenG5}, we 
consider the one-loop dressed 
approximation of $G_5^{\mu\nu}$, and show that the WTI is exactly fulfilled provided 
that the quark-gluon vertex satisfies its own SDE, also approximated at the 
one-loop dressed level. 
Then, the self-consistent implementation of the 
above-mentioned approximations at the level of the 
SDE for $\ga^{\mu}$ is carried out 
in \sect{sec:sdetrun}. In \sect{sec:Disc} we 
summarize our results and discuss possible future directions.
Lastly, in \appref{app:WTIproof} we 
demonstrate the WTIs obeyed by  
$\ga^{\mu}$ and $G_5^{\mu\nu}$, while in \appref{app:singletnonsinglet} we comment on the differences between flavour singlet and non-singlet axial-vector currents.

%%%%%%%%%%%%%%%%%%%%%%%%%%%%%%%%
\section{Quark propagator and quark-gluon vertex}\label{sec:GapEq}

In this section we review the 
main properties of two fundamental ingredients of our analysis, 
namely the quark propagator and the 
quark-gluon vertex. 

The starting point of our considerations is the quark propagator, $S^{ab}(p)$,
depicted in \fig{fig:Elementary}, 
which we cast 
in the standard form
$S^{ab}(p)=i\delta^{ab}S(p)$,
see 
 \eg\cite{Itzykson:1980rh}.
Typically, one decomposes the inverse 
quark propagator, $S^{-1}(p)$, as 
\be\label{eq:InvS}
    S^{-1}(p)=A(p^2)\slashed{p}-B(p^2)~,
\ee
where $A(p^2)$ and $B(p^2)$ are the dressings of the (Dirac) vector and scalar structures,  respectively. The renormalization-group invariant (RGI) quark mass function, 
${\mathcal M}(p^2)$,  
is given by \mbox{${\mathcal M}(p^2) = B(p^2)/ A(p^2)$}. 

For the purposes of the present analysis, it is useful to also introduce the Dirac decomposition of $S(p)$, namely 
\be
S(p)=a(p^2)\slashed{p}+b(p^2) \,,
\label{eq:S}
\ee
with 
\be 
a(p^2)=c(p^2)A(p^2) \,,\qquad
b(p^2)=c(p^2)B(p^2) \,,\qquad
c(p^2):=\frac{1}{A^2(p^2)p^2-B^2(p^2)}
\,.
\label{eq:abc}
\ee

\begin{figure}[t!]
    \centering
    \includegraphics[scale=1]{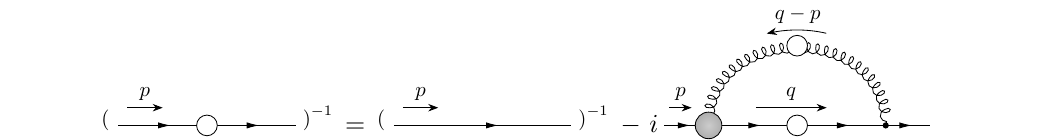}
    \caption{Pictorial representation of the quark gap equation. The white circles denote 
    full propagators, while the gray circle 
    stands for the fully-dressed 
    quark-gluon vertex, $\Gamma_{\mu}$.}
    \label{fig:GapEqPict}
\end{figure}

The momentum evolution of the functions 
$A(p^2)$  and $B(p^2)$ is determined from  
the SDE that governs the quark propagator, the \textit{gap equation},
\begin{align}
\label{eq:GapEq}
\is(p)= \slashed{p} - m -i\Sigma(p)\,,
\end{align}
and it is shown diagrammatically in \fig{fig:GapEqPict}.
In \1eq{eq:GapEq}, $\Sigma(p)$
stands for the 
quark self-energy, 
\begin{align}
\label{eq:SeflEnergy}
    \Sigma(p)=-g^2C_f\int_q\gamma^\nu S(q)\g^\mu(q-p,p,-q)\Delta_{\mu\nu}(q-p)\,,
\end{align}
where $g$ is 
the QCD gauge coupling and $C_f=4/3$ the Casimir eigenvalue of the fundamental SU(3) representation.
In addition, 
the integral measure is denoted by  
\begin{align}
\int_q :=  \int_{R^4}\frac{{\rm d}^4 q}{(2\pi)^4} \,,
\end{align}
where the use of a symmetry-preserving regularization scheme is implicitly assumed.
Moreover, 
$\Delta_{\mu\nu}$ and $\g_\mu$  represent the full gluon propagator and the fully-dressed quark-gluon vertex, respectively, to be discussed below.
Finally, $m$ stands for the 
current quark mass; 
note that, 
throughout this work, we restrict 
ourselves to 
the case $m=0$ (chiral limit).

In the covariant ($R_{\xi}$) gauges \cite{Fujikawa:1972fe}, the gluon propagator,  \mbox{$\Delta^{ab}_{\mu\nu}(q)=-i\delta^{ab}\Delta_{\mu\nu}(q)$}, assumes the general form 
\begin{align}
    \Delta_{\mu\nu}(q) = \Delta(q^2)\,P_{\mu\nu}(q) + \xi \frac{q_{\mu}q_{\nu}}{q^4}\,,
    \label{eq:gluoprop}
\end{align}
where $P_{\mu\nu}(q) = g_{\mu\nu} - 
q_{\mu}q_{\nu}/q^2$ is the transverse projection operator and 
$\xi$ is the gauge-fixing parameter;
for the diagrammatic representation
of the gluon propagator see \fig{fig:Elementary}. The function 
$\Delta(q^2)$ denotes the scalar component of the gluon
propagator and depends explicitly on $\xi$. Note that in practical applications one uses 
almost exclusively the Landau gauge 
$(\xi=0)$, which brings about certain simplifications in the treatment of the gap equation. 
However, in the present analysis we will keep 
a general value of $\xi$, 
since, as we will see,
none of the ensuing demonstrations  
depends on the gauge choice. 

The quark-gluon vertex (see \fig{fig:Elementary}) is defined as \mbox{$\g^{b}_{\!\mu}(q,r,-p)=ig\frac{\lambda^b}{2}\g_{\!\mu}(q,r,-p)$}, with $\lambda^b$ denoting 
the standard Gell-Mann matrices
($b=1,...,8$).
The part $\g^\mu(q,r,-p)$
is typically decomposed in a tensorial basis composed by 12 elements, to be denoted by 
$\tau_i^\mu(q,r,-p)$; for some standard choices of bases in the 
literature, see, \eg \cite{Kizilersu:1995iz,Davydychev:2000rt,Aguilar:2014lha,Binosi:2016wcx}.
Thus, 
\begin{align}
\g^\mu(q,r,-p)=\sum_{i=1}^{12}\lambda_i(q,r,-p)\tau_i^\mu(q,r,-p)\,,
\label{qgtens}
\end{align}
where the scalar functions 
$\lambda_i(q,r,-p)$ are the 
associated form factors or 
dressings. 
Note that, in the Landau gauge ($\xi=0$), 
the number of basis elements is reduced to eight, and the vertex 
is usually referred to as the ``transversely projected" vertex, see, \eg \cite{Gao:2021wun}, 
and references therein. 

In what follows we will make extensive use of the separation 
of $\g^\mu$ into two components,
$\go^\mu$ and $\ge^\mu$, 
comprised by the basis elements 
$\tau_i^\mu$ that contain an odd or even number of Dirac $\gamma$ matrices,
respectively. 
In particular, we may enumerate 
the basis elements $\tau_i^\mu$ such that the first (last) six 
contain an odd (even) number of Dirac
$\gamma$ matrices,
and define 
\begin{align}
\go^\mu(q,r,-p) :=&\, \sum_{i=1}^{6}\lambda_i(q,r,-p)\tau_i^\mu(q,r,-p)\,,
\nonumber\\[1ex]
\ge^\mu(q,r,-p) :=&\, \sum_{i=7}^{12}\lambda_i(q,r,-p)\tau_i^\mu(q,r,-p)\,,
\label{eq:tauOddEven}
\end{align}
such that   
\begin{align}
\g^\mu(q,r,-p)=
\underbrace{\go^\mu(q,r,-p)}_{{\rm odd\,\#\,of\,\gamma}} \,\,+ \,\, \underbrace{\ge^\mu(q,r,-p)}_{{\rm even\,\#\,of\,\gamma}}  \,.
\label{eq:QgOddEven}
\end{align}
For a given basis, the above decomposition is  unique. As it will become clear in what follows, 
the decomposition in \1eq{eq:QgOddEven}
appears naturally in various points of the 
subsequent analysis, where we will take advantage of the elementary relations 
\begin{align}
\textrm{Tr}\!\left[(\rm even\,\#\, of \,\gamma) \times \go^\mu \right] =0 =\textrm{Tr}\!
\left[(\rm odd\,\#\, of \,\gamma) \times \ge^\mu\right] \,, 
\end{align}
and, in addition, 
\begin{align}
\gamma_5 \g^\mu = 
\gamma_5 (\go^\mu +\ge^\mu)
= (\ge^\mu - \go^\mu)\gamma_5
\,.
\label{eq:QgOddEvengamma5}
\end{align}

\begin{figure}[t!]
    \hspace*{-0.25cm}
    \includegraphics[scale=1]{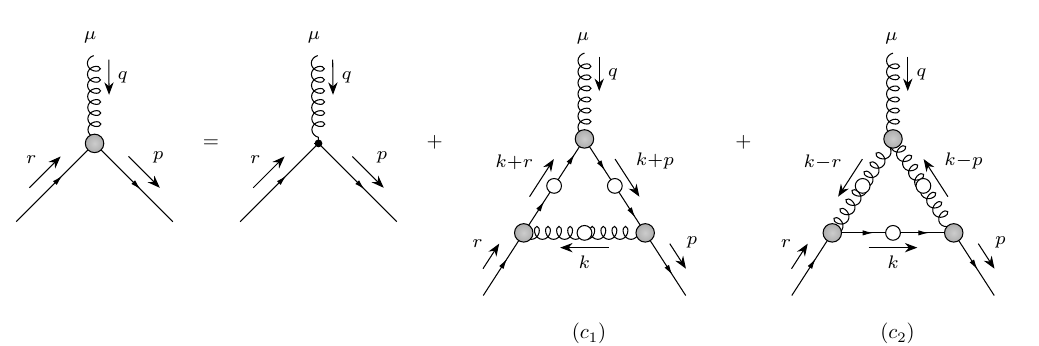}
    \caption{Diagrammatic representation of the full quark-gluon vertex SDE,  obtained from a three-loop truncation of the 3PI effective action. Gray circles represent the fully-dressed quark-gluon and three-gluon vertices, while white ones identify full quark and gluon propagators. Diagrams $(c_1^\mu)$ and $(c_2^\mu)$ are often referred to as ``abelian'' and ``non-abelian'', respectively.}
    \label{fig:QGSDE}
\end{figure}

In principle, the evolution of the 
form factors $\lambda_i$
is determined from the full SDE satisfied 
by $\g_\mu(q,r,-p)$.
To be sure, in practice one 
employs truncated versions 
of the SDE, thus obtaining approximate results for the $\lambda_i$. 
One of the standard approximations of the SDE, extensively employed in the recent literature, 
is shown in \fig{fig:QGSDE}; 
this particular SDE is obtained from the 3PI effective action  
at the three-loop order~\cite{Berges:2004pu,Berges:2004yj,York:2012ib,Carrington:2010qq,Alkofer:2008tt,Williams:2015cvx}. 
One of the special characteristics of the SDE in \fig{fig:QGSDE}
is that all 
fundamental vertices inside the diagrams $(c_1^\mu)$ and $(c_2^\mu)$ are fully dressed. 
This is to be contrasted with the standard 
SDE formulation (see \eg \cite{Alkofer:2008tt} and references therein), where one of the vertices is always kept at its classical (tree-level) form. Note that, equivalently, one may reach 
the form of the SDE in \fig{fig:QGSDE} by resorting to  the standard skeleton expansion~\cite{Gao:2024gdj}.

The dynamical equation depicted in \fig{fig:QGSDE} 
represents a typical case of an SDE truncated at 
the {\it ``one-loop dressed''} level: it is comprised by the pair of one-loop diagrams known from perturbation theory, but with all propagators and vertices fully dressed. 
Given that this particular version of the SDE  plays a crucial rôle in 
the construction presented in 
\sect{sec:OpenG5}, 
we find it useful to denote the corresponding solution 
by $\g_\mu^{({\bf \s 1})}$, namely 
\begin{align}
\g_\mu^{({\bf \s 1})}=\gamma_\mu+(c_{1\mu})+(c_{2\mu})\,,
\label{Goneup}
\end{align}
with the two loop parts $c_1$ and $c_2$. They read  
\begin{align} \nonumber 
(c_{1\mu}) = &\, c_a\!\int_k\g_\beta^{({\bf \s 1})}(-k,k+p,-p)S(k+p)\g_\mu^{({\bf \s 1})}(q,k+r,-k-p)\\[1ex]
&\, \hspace{.5cm} \times S(k+r)\g_\alpha^{({\bf \s 1})}(k,r,-k-r)\Delta^{\alpha\beta}(k)\,,
\end{align} 
and 
\begin{align}\nonumber 
(c_{2\mu}) = &\, c_b\!\int_k \g_\beta^{({\bf \s 1})}(p-k,k,-p)S(k)\g_\alpha^{({\bf \s 1})}(k-r,r,-k)\g_{\!\mu\rho\sigma}(q,k-p,k-r)\\[1ex]
&\, \hspace{.5cm} \times \Delta^{\rho\beta}(k-p)\Delta^{\alpha\sigma}(k-r)\,, 
\end{align}
where $\g_{\!\mu\rho\sigma}$ is the full three-gluon vertex shown in \fig{fig:Elementary} and with the prefactors 
\begin{align}
c_a:=-ig^2\left(C_f-\frac{C_A}{2}\right) \,, 
\qquad\qquad c_b:=ig^2 \frac{C_A}{2} \,.
\label{cacb}
\end{align}
We emphasize once again 
that $\g_\mu$ is the exact quark-gluon vertex, while 
$\g_\mu^{({\bf \s 1})}$ is an approximation to it, obtained from 
a truncated SDE. Thus, in general, 
$\g_\mu \neq \g_\mu^{({\bf \s 1})}$, 
even though the differences are 
expected to be small. The distinction between $\g_\mu$ 
and $\g_\mu^{({\bf \s 1})}$ will be  maintained in what follows; thus, 
the exact results presented in 
Secs.~\ref{sec:WTI}, \ref{sec:SDE}, 
and \ref{sec:ChiralLim}
involve
the former, while the 
one-loop dressed 
construction of \sect{sec:OpenG5} entails the latter. 

Returning to the quark gap equation
of \1eq{eq:GapEq}, 
we may suitably project out the 
coupled system that determines 
$A(p^2)$ and $B(p^2)$; specifically,
taking the trace of \1eq{eq:GapEq}
isolates 
the equation for $B(p^2)$, while 
multiplying by $\slashed{p}$ and then taking the trace selects $A(p^2)$. Thus, we obtain 
\bea\label{eq:AEq}
    p^2 A(p^2) = p^2 & + & \frac{ig^2 C_f}{4}\int_q~a(q^2)\textrm{Tr}\left[\slashed{p}\gamma^\nu\slashed{q}\go^\mu(q-p,p,-q)\right]\Delta_{\mu\nu}(q-p)\nonumber\\
    \nonumber\\
     & + & \frac{ig^2 C_f}{4}\int_q~b(q^2)\textrm{Tr}\left[\slashed{p}\gamma^\nu\ge^\mu(q-p,p,-q)\right]\Delta_{\mu\nu}(q-p)\,,
\eea
and
\bea\label{eq:BEq}
    B(p^2) =& - &\displaystyle\frac{ig^2 C_f}{4}\int_q~a(q^2)\textrm{Tr}\left[\gamma^\nu\slashed{q}\ge^\mu(q-p,p,-q)\right]\Delta_{\mu\nu}(q-p)\nonumber\\
    \nonumber\\
    &-& \frac{ig^2 C_f}{4}\int_q~ b(q^2)\textrm{Tr}\left[\gamma^\nu\go^\mu(q-p,p,-q)\right]\Delta_{\mu\nu}(q-p)\,,
\eea
where we 
note in both  \1eq{eq:AEq} and 
\1eq{eq:BEq} 
the appearance of the 
components $\go^\mu$ and $\ge^\mu$,
introduced in \1eq{eq:QgOddEven}.

Observe finally that in the RL approximation, where $\g_\mu \to \gamma_\mu$ (and $\xi=0$), the quark self-energy of \1eq{eq:SeflEnergy} becomes 
\begin{align}
\label{eq:SERL}
    \Sigma_{\rm {\s R \s L}}(p)=-g^2C_f\int_q\gamma^\nu S(q)\gamma^\mu \Delta_{\mu\nu}(q-p)\,,
\end{align}
and the system of \2eqs{eq:AEq}{eq:BEq}
collapses to 
\begin{align}\nonumber 
p^2 A(p^2)  = &\, p^2 + \frac{ig^2 C_f}{4}\int_q~a(q^2)\textrm{Tr}\left[\slashed{p}\gamma^\nu\slashed{q}\gamma^\mu\right]\Delta_{\mu\nu}(q-p) \,,\\[1ex]
B(p^2) = &\,
    -\frac{ig^2 C_f}{4}\int_q~ b(q^2)\textrm{Tr}\left[\gamma^\nu\gamma^\mu\right]\Delta_{\mu\nu}(q-p)\,.    
\label{eq:ABRL}
\end{align}
%

%%%%%%%%%%%%%%%%%%%%%%%%%%%%%%%%%%%
\section{Axial vertices and their Ward-Takahashi identities}\label{sec:WTI}

\begin{figure}[t!]
    \hspace*{-0.75cm}
    \includegraphics[scale=1]{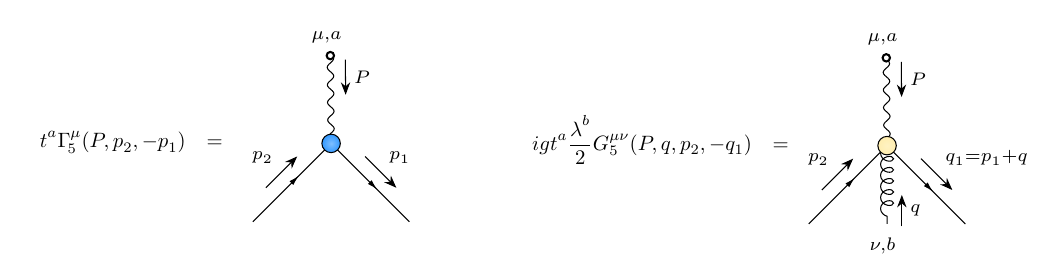}
    \caption{Diagrammatic representation of the two central vertices treated in this work: \textit{left-panel}: axial-vector vertex; \textit{right-panel}: gluon-axial-vector vertex.}
    \label{fig:Elementary5}
\end{figure}

When contracted by $P_{\mu}$, 
the axial vertices 
$\ga^{\mu}(P, p_2,-p_1)$ and 
$G_5^{\mu\nu}(P,p_2,q,-p_1-q)$, shown in 
\fig{fig:Elementary5},
satisfy abelian WTIs,
denoted by \wtig and \wtiG, 
respectively. While 
the former is well-known, see, \eg \cite{Fujikawa:1980eg,Miransky:1994vk}, the latter has received little attention \cite{Bender:2002as,Bhagwat:2004hn,Chang:2009zb}.
In this section we discuss these 
WTIs, and the constraints they impose on the pole structure of the two axial vertices.

%

%%%%%%%%%%%%%%%%%%%%%%%%%%%%%%%%%%%%
\subsection{Axial-vector vertex \texorpdfstring{$\ga^\mu$}{g5}}
\label{subsec:av}

Consider the flavour non-singlet axial-vector vertex 
$\ga^\mu$ (\fig{fig:Elementary5}), 
defined as the amputated part of the correlation function  
$\langle 0|Tj^{\mu}_5(y)\psi(x_1)\bar{\psi}(x_2)|0\rangle$ (see \appref{app:WTIproof}). The (non-singlet) axial-vector current is given by \mbox{$j^{\mu}_{5}(x) = \bar{\psi}(x)\gamma_5 \gamma^\mu t^a{\psi}(x)$}, where the $t^a$ are the flavour $SU(N_f)$ generators which, for the shortness, we are systematically dropping.

In momentum space, and in the absence of current quark masses 
($m=0$), the vertex
$\ga^\mu(P,p_2,-p_1)$ satisfies the axial WTI \cite{Fujikawa:1980eg,Itzykson:1980rh,Miransky:1994vk} (see \appref{app:WTIproof} for the derivation)
\be
\label{eq:WTI5}
[{\rm WTI}]_{\Gamma_{\!5}}~:\qquad -P_\mu\ga^\mu(P,p_2,-p_1)=\is(p_1)\gamma_5+\gamma_5\is(p_2)\,.
\ee

In the limit $P\to 0$, in which case $p_1=p_2 :=p$, this identity reveals 
a profound connection between the 
term $B(p^2)$, which 
emerges when the chiral symmetry is dynamically broken, and the  
amplitude that controls the formation of a pion as a quark-antiquark bound state. 

In particular, it is well-known that, in the presence of a nonvanishing $B(p^2)$, the only way to 
satisfy 
\1eq{eq:WTI5} is to allow 
$\ga^\mu(P,p_2,-p_1)$ to exhibit 
a pole at $P^2=0$, which is associated 
with the corresponding massless Goldstone boson,
\ie the pion. 

Specifically, if we substitute 
the decomposition of \1eq{eq:InvS} 
into \1eq{eq:WTI5} and implement the 
limit $P\to 0$, the terms proportional to $A(p^2)$ drop out, and we find
\be
\label{eq:WTIP0-I}
\lim_{P\to 0}P_\mu\ga^\mu(P,p_2,-p_1)=2B(p^2)\gamma_5\,.
\ee
This observation leads us to conclude that $\ga^{\mu}$ must display a pole in $P^2$, \ie
\be\label{eq:g5Decomp}
\ga^\mu(P,p_2,-p_1)=\left.\ga^\mu (P,p_2,-p_1)\right|_{\textrm{pole}}+\left.\ga^\mu(P,p_2,-p_1)\right|_{\textrm{reg}}~,
\ee
with
\be\label{eq:g5Pole}
\left.\ga^\mu(P,p_2,-p_1)\right|_{\textrm{pole}}=\frac{P^\mu}{P^2} \,\chi(P,p_2,-p_1)\gamma_5~,
\ee
where $\chi(P,p_2,-p_1)$ is the BSA of the pion. Note that 
we suppress the pion decay constant, $f_\pi$, which is typically  introduced at this point, 
see, \eg\cite{Maris:1997hd,Miransky:1994vk}. 

\begin{figure}[!t]
    \hspace*{-0.6cm}
    \includegraphics[scale=1]{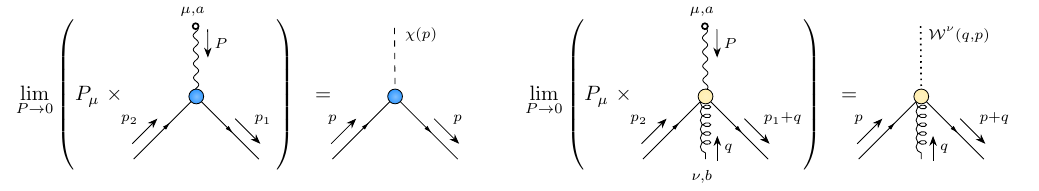}
    \caption{Diagrammatic representation of the pole contributions to the two axial vertices under consideration: {\it left-panel}: pole residue of the axial-vector vertex; {\it right-panel}: pole residue of the gluon-axial-vector vertex.}
    \label{fig:P0axialvertices}
\end{figure}

Using \1eq{eq:g5Decomp}, it is straightforward to show that, in the limit $P\to 0$, 
\be
\label{eq:WTIP0-II}
\lim_{P\to 0}P_\mu\ga^\mu(P,p_2,-p_1)=\lim_{P\to 0}\left. P_\mu\ga^\mu(P,p_2,-p_1)\right|_{\textrm{pole}}=\chi(0,p,-p)\gamma_5 \,.
\ee
a result whose diagrammatic representation is shown in the left-panel of \fig{fig:P0axialvertices}. Then, combining 
\2eqs{eq:WTIP0-I}{eq:WTIP0-II}, 
we find 
\be
\chi(0,p,-p)=2B(p^2) \,,
\label{eq:WTIP0-3}
\ee
where the $\gamma_5$ has been canceled 
from both sides. 

The relation in \1eq{eq:WTIP0-3}
may be further inspected, by resorting to the standard tensorial decomposition for the amplitude 
$\chi(P,p_2,-p_1)$~\cite{Miransky:1994vk}
\be\label{eq:chi}
\chi(P,p_2,-p_1)=\chi_1+\chi_2\slashed{P}+\frac{1}{2}\chi_3(\slashed{p}_1+\slashed{p}_2)+i\chi_4[\slashed{p}_1,\slashed{p}_2]~,
\ee
where the $\chi_i = \chi_i(P,p_2,-p_1)$ are the associated form factors,
which depend on three Lorentz scalars (\eg $p_1^2$, $p^2_2$ and $p_1\cdot p_2$). In particular, when $P \to 0$, we have 
\begin{align}
\chi(0,p,-p) = \chi_1(p^2)
+ \chi_3(p^2)\slashed{p} \,.
\label{chi0pp}
\end{align}
Then, from \1eq{eq:WTIP0-3} we obtain 
the two basic relations \cite{Maris:1997hd,Fischer:2008wy},
\begin{align}
\chi_1(p^2)=2B(p^2) \,,
\label{eq:chi1chi1}
\end{align}
and 
\begin{align}
\chi_3(p^{2})=0 \,.
\label{eq:chi1chi3}
\end{align}
Notably, \1eq{eq:chi1chi3} may also be directly derived from the tensor decomposition in \1eq{eq:chi}, as a consequence of charge conjugation invariance.

%%%%%%%%%%%%%%%%%%%%%%%%%%%%%%%%%%%%
\subsection{Gluon-axial-vector vertex \texorpdfstring{$G_5^{\mu\nu}$}{G5}}
\label{subsec:gav}

The second main element of our construction is the gluon-axial-vector vertex, $G_5^{\mu\nu}$ (\fig{fig:Elementary5}). This particular vertex is defined as the 
amputated and \opi part of the 
correlation function comprised by 
a quark-antiquark 
pair, a gluon field, and an axial-vector current $j_5^\mu$, 
\appref{Appb}.

In the absence of current quark masses, $G_5^{\mu\nu}(P,q,p_2,-q_1)$ satisfies the axial WTI (see \appref{app:WTIproof} for the derivation and \cite{Eichmann:2016yit, Chang:2009zb, Bender:2002as, Bhagwat:2004hn} for additional discussions).
\begin{align}
[{\rm WTI}]_{G_5}~:\qquad -iP_\mu G_5^{\mu\nu}(P,q,p_2,-q_1)=\g^\nu(q,p_1,-q_1)\gamma_5+\gamma_5\g^\nu(q,p_2,-q_2)\,,
\label{eq:G5WTI}
\end{align}
where $q_i:=p_i+q$. Note that, since 
$\gamma^{\nu} \gamma_5 +
\gamma_5 \gamma^{\nu} =0$,
the combination of vertices 
appearing on the r.h.s. of \1eq{eq:G5WTI}
does not contain tree-level contributions, \ie 
\begin{align}
    -iP_\mu G_5^{\mu\nu}(P,q,p_2,-q_1)=\g^\nu_{\!\s Q} (q,p_1,-q_1)\gamma_5+\gamma_5 \g^\nu_{\!\s Q}(q,p_2,-q_2) \,,
\label{eq:G5WTInotree}
\end{align}
where
\begin{align}
\g^\nu_{\!\s Q}(q,r,-p):=\g^\nu(q,r,-p)-\lambda_1(q,r,-p)\gamma^\nu\,.
\label{eq:GQ}
\end{align}
This observation is consistent with the fact that 
the vertex $G_5^{\mu\nu}$, 
appearing on the l.h.s. 
of \1eq{eq:G5WTInotree}, 
does not possess a tree-level term.

As in the case of the axial-vector vertex, this WTI imposes strong constraints on $G_{5}^{\mu\nu}$. Similarly to the previous case, \1eq{eq:G5WTI} requires the gluon-axial-vector vertex to have a longitudinally coupled pole. In fact, taking the limit $P\rightarrow 0$ of this identity, the terms involving $\go^\nu$ drop out, and we are left with
\begin{align}\nonumber 
\lim_{P\to 0}P_\mu G_5^{\mu\nu}(P,q,p_2,-q_1)  = &\, i\left[\g^\nu(q,p,-p-q)\gamma_5+\gamma_5\g^\nu(q,p,-p-q)\right]\\[1ex]
 = &\, 2i\ge^\nu(q,p,-p-q)\gamma_5\,,
\label{eq:G5poleG2}
\end{align}
where
$\ge^\nu(q,p,-p-q)$
is the even part of the 
quark-gluon vertex
$\g^\nu(q,p,-p-q)$.
Note that $\g^\nu(q,p,-p-q)$ corresponds to the quark-gluon vertex shown in \fig{fig:QGSDE}, 
but with 
$r \leftrightarrow p$.  

We emphasize that, crucially, 
$\ge^\nu$ is comprised precisely 
by the terms  referred to as 
\textit{chiral symmetry breaking} components (see, \eg \cite{Gao:2021wun,Aguilar:2024ciu}):  when solving the SDE 
of the quark-gluon vertex, they arise only when 
a nontrivial $B(p^2)$ has emerged 
from the quark gap equation. 
Thus, it is precisely the 
part of the quark-gluon vertex  
induced by the breaking of chiral symmetry that saturates  
the pole residue of 
$G^{\mu\nu}_5$. 

To explore this point in some detail, 
we note that, in complete analogy to 
$\Gamma_{5}^{\mu}$, 
\1eq{eq:G5poleG2} can be resolved only if $G^{\mu\nu}_5$ possesses 
a pole 
in $P^2=0$, \ie 
\begin{align}
G^{\mu\nu}_5 (P,q,p_2,-q_1)=\left.G^{\mu\nu}_5(P,q,p_2,-q_1)\right|_{\textrm{pole}}+\left.G^{\mu\nu}_5(P,q,p_2,-q_1)\right|_{\textrm{reg}}\,,
\label{eq:G5decomposition}
\end{align}
with
\begin{align}
\left.G^{\mu\nu}_5(P,q,p_2,-q_1)\right|_{\textrm{pole}}=2i\,\frac{P^\mu}{P^{2}}\,\mathcal{W}^\nu(P,q,p_2,-q_1)\gamma_5\,.
\label{eq:G5pole}
\end{align}
At the formal level, 
the component 
$\mathcal{W}^\nu$ plays a rôle analogous to that of the amplitude $\chi$ in \1eq{eq:g5Pole}, acting as a momentum-dependent residue.
From the physical point of view, 
$\mathcal{W}^\nu$
encapsulates key  pieces of the pion dynamics, contained
in the SDE of the
axial-vector vertex.
In that sense, it may be viewed as a  
concise representation 
of various contributions 
that participate in the momentum evolution of the pion amplitude 
$\chi$.

Using the decomposition of \1eq{eq:G5decomposition} 
in conjunction with \1eq{eq:G5pole}, it is straightforward to show that  (\fig{fig:P0axialvertices}, right-panel)
\begin{align}\nonumber 
\lim_{P\to 0}P_\mu G_5^{\mu\nu}(P,q,p_2,-q_1) = &\, \lim_{P\to 0}P_\mu\left.G_5^{\mu\nu}(P,q,p_2,-q_1)\right|_{\textrm{pole}}\\[1ex]
= &\, 2i\,\mathcal{W}^\nu(0,q,p,-p-q)\gamma_5=2i\,\ge^\nu(q,p,-p-q)\gamma_5\,,
\label{eq:G5poleContr}
\end{align}
where, once again, $\ge^\nu(q,p,-p-q)$
is the even part of the 
quark-gluon vertex
$\g^\nu(q,p,-p-q)$.

It is clear that $\mathcal{W}^\nu(0,q,p,-p-q)$ 
admits the same tensorial decomposition as the quark-gluon vertex $\g^\nu(q,r,-r-q)$ in \1eq{qgtens}, with the change $r\leftrightarrow p$, namely 
\begin{align}
\mathcal{W}^\nu(0,q,p,-p-q) 
=\sum_{i=1}^{12} w_i(q,p,-p-q)\tau_i^\nu(q,p,-p-q)\,,
\label{Wtens}
\end{align}
where the $w_i(q,p,-p-q)$
are the corresponding 
form factors (dressings). 
Evidently, 
as in \1eq{eq:QgOddEven},
a separation 
into odd and even structures 
may be carried out
(suppressing momenta)
\begin{align}
\mathcal{W}^\nu 
= \mathcal{W}_1^{\nu} +
\mathcal{W}_2^{\nu} \,,
\end{align}
with 
\begin{align}
\mathcal{W}_1^{\nu} =
\sum_{i=1}^{6} w_i\tau_i^\nu
\,,
\qquad
\mathcal{W}_2^{\nu} =
\sum_{i=7}^{12} w_i\tau_i^\nu
\,.
\end{align}
The substitution of the above 
tensor decomposition into \1eq{eq:G5poleContr}
leads directly to the relations
\begin{align}
\mathcal{W}_2^{\nu}=\ge^\nu(q,p,-p-q) \,,
\qquad\qquad
\mathcal{W}_1^{\nu} = 0 , 
\end{align}
or, equivalently, at the 
level of the form factors 
\begin{align}
w_i(q,p,-p-q) = \lambda_i (q,p,-p-q) \,, \qquad i=7,...,12 \,,
\end{align}
and 
\begin{align}
w_i(q,p,-p-q) = 0  \,, \qquad i=1,...,6 \,;
\end{align}
which represent the direct analogues of \2eqs{eq:chi1chi1}{eq:chi1chi3}, respectively.

%%%%%%%%%%%%%%%%%%%%%%%%%%%%%%%%%%%%%
\section{The axial WTI from the vertex SDE: exact derivation}
\label{sec:SDE}

\begin{figure}[t!]
    \centering
    \includegraphics[scale=1]{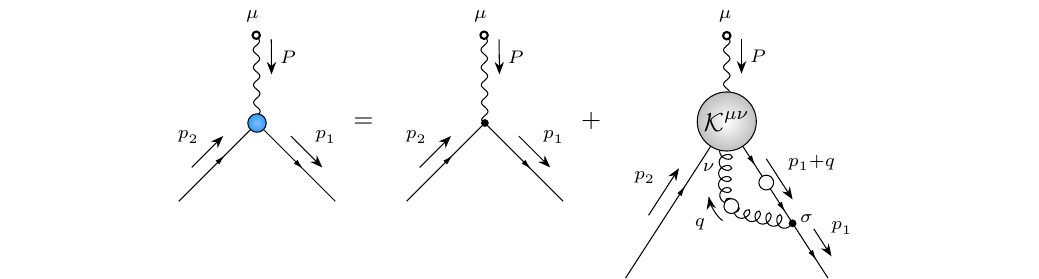}
    \caption{Diagrammatic representation of the SDE satisfied by the axial-vector vertex $\ga^\mu$, 
    formulated from the point of view of the antiquark leg.}
    \label{fig:kernel}
\end{figure}

In this section we demonstrate that 
the exact dynamical equation (SDE)
that governs the evolution of $\ga^\mu$
satisfies precisely the 
\wtig of \1eq{eq:WTI5}.

To that end, instead of considering the 
SDE satisfied by 
$\ga^\mu$ from the perspective of the axial-vector current (leg with momentum $P$),  
we adopt the view 
of the antiquark leg, carrying 
momentum $p_1$;
the resulting SDE is shown in \fig{fig:kernel}. 
According to the standard rules, 
the vertex connecting the 
special leg to the 
rest of the diagram 
must be kept at tree-level
($\gamma^{\sigma}$ in \fig{fig:kernel}). 
This particular way of  writing the  
SDE, \ie using an  antifermion as the reference leg
has been used in the studies of the 
ghost-gluon~\cite{Fischer:2006ub,Aguilar:2013xqa,Ferreira:2025anh} 
and quark-gluon 
vertices~\cite{Binosi:2016rxz}. 

\begin{figure}[t!]
\centering
\includegraphics[scale=1]{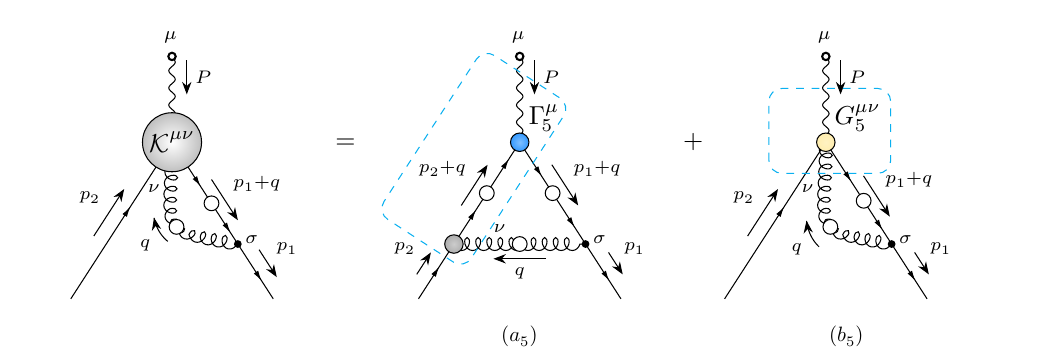}
\caption{
Diagrammatic representation of the 
quantum part of the SDE in \fig{fig:kernel}.
The decomposition of $\mathcal{K}^{\mu\nu}$ given in \1eq{eq:kernel1pr1pi} is 
delimited by the two blue-dashed boxes.}
\label{fig:SDEPicture}
\end{figure}

Thus, the resulting SDE reads 
\begin{align}
\ga^\mu(P,p_2,-p_1) = \gamma_5\gamma^\mu+ig^2C_f\int_q \gamma^\sigma S(q_1)\mathcal{K}^{\mu\nu}(P,q,p_2,-q_1)\Delta_{\nu\sigma}(q)\,,
\label{eq:SDE}
\end{align}
where we have defined $q_1 :=p_1+q$.
The kernel $\mathcal{K}^{\mu\nu}$
is comprised by four incoming 
operators, a quark ($\psi$), an antiquark ($\bar{\psi}$), a gluon ($A^\nu$), and a axial-vector current ($j_5^\mu$), 
so that 
\mbox{$\mathcal{K}^{\mu\nu} :=
\mathcal{K}_{j_5^\mu\! A^\nu\! \psi\bar{\psi}}$}. This kernel may be decomposed as shown in \fig{fig:SDEPicture}, namely 
\begin{align}
\mathcal{K}^{\mu\nu}(P,q,p_2,-q_1)=-\ga^\mu(P,q_2,-q_1)S(q_2)\g^\nu(q,p_2,-q_2)+iG_5^{\mu\nu}(P,q,p_2,-q_1)\,,
\label{eq:kernel1pr1pi}
\end{align}
with $q_2 :=p_2+q$, and  
$G_5^{\mu\nu}$ is the  
gluon-axial-vector vertex 
introduced in Sec.~\ref{subsec:gav}.

As a result, 
the SDE of $\ga^\mu$ can be written as 
\begin{align}
\ga^\mu =\gamma_5\gamma^\mu+ (a^\mu_5) +(b^\mu_5)\,,
\label{eq:SDEav}
\end{align}
where
\begin{align}
     (a_5^\mu) = &\, -ig^2C_f\int_q \gamma^\sigma S(q_1)\ga^\mu (P,q_2,-q_1)S(q_2)\g^\nu (q,p_2,-q_2)\Delta_{\nu\sigma}(q)~,
     \label{eq:a5mu}
     \\[1ex]
     (b^\mu_5)  = &\, -g^2C_f\int_q \gamma^\sigma S(q_1)G_5^{\mu\nu}(P,q,p_2,-q_1)\Delta_{\nu\sigma}(q)\,.
     \label{eq:b5mu}
\end{align}
Note that the 
lowest-order diagram contributing to  
$\mathcal{K}^{\mu\nu}$ is 
enclosed in the blue-dashed box of 
$(a_5^\mu)$ in \fig{fig:SDEPicture}.  
A ``cut" through the quark propagator with momentum $p_2+q$ shows that 
this diagram is clearly one-particle reducible (1PR); of course,  
the full diagram, 
$(a_5^\mu)$, where the kernel is embedded, is one-particle irreducible (1PI).  
All remaining graphs contained in 
$\mathcal{K}^{\mu\nu}$ are \opi, 
and coincide exactly with the graphs 
forming the vertex $G_5^{\mu\nu}$, which is enclosed entirely within the 
blue-dashed box
in graph $(b_5^\mu)$. 

If, instead, a flavour singlet axial-vector current is considered, a third 
diagrammatic contribution 
is present in the decomposition of $\mathcal{K}$, 
coupling the axial-vector current to a pair of gluons , see \eg \cite{Eichmann:2016yit} and references therein. However, as explained in \appref{app:singletnonsinglet}, this additional 
term vanishes in the non-singlet case, while in the 
singlet case contributes solely to the anomaly. 
In that sense, as long as the SDE of \1eq{eq:SDEav} describes the flavour non-singlet vertex, it is exact.

Given that the graphs 
$(a^\mu_5)$ and $(b^\mu_5)$ account for the full 
sum of all 
quantum corrections, if we contract 
both sides of \1eq{eq:SDEav} by $P_\mu$,
\begin{align}
P_\mu\ga^\mu(P,p_2,-p_1)=- (\sla{p_1}\gamma_5 +  
\gamma_5 \sla{p_2})
+P_\mu \left[(a_5^\mu) + 
(b_5^\mu)\right]\,, 
\label{eq:SDEWTI-I}
\end{align}
the r.h.s. of the \wtig in \1eq{eq:WTI5}  must emerge exactly. 

Let us focus on the contraction of graph $(a_5^\mu)$, 
which is the only loop contribution 
typically considered within the RL  approximation. Using the r.h.s of 
\1eq{eq:WTI5} for
the term $P_{\mu}\ga^\mu (P,q_2,-q_1)$ under the integral sign, 
we find 
\begin{align} \nonumber
    P_\mu(a_5^\mu) & = ig^2C_f\int_q\gamma^\sigma S(q_1)\left[\is(q_1)\gamma_5+\gamma_5\is(q_2)\right]S(q_2)\g^\nu(q,p_2,-q_2)\Delta_{\nu\sigma}(q)\\[1ex]\nonumber 
    = &\, \gamma_5\underbrace{\left[-ig^2 C_f\int_q\gamma^\sigma S(q_2)\g^\nu(q,p_2,-q_2)\Delta_{\nu\sigma}(q)\right]}_{i\Sigma(p_2)}\\[1ex]
    &\, + \underbrace{ig^2 C_f\int_q\gamma^\sigma S(q_1)\gamma_5\g^\nu(q,p_2,-q_2)\Delta_{\nu\sigma}(q)}_{V(p_1,p_2):\,{\textrm{symmetry violating term}}}\,.
 \label{Pa5} 
 \end{align}
The result in \1eq{Pa5} 
prompts the following observations: %
\begin{enumerate}[label=(\textit{\roman{enumi}})]
\item When 
the term  $i \Sigma(p_2)$
of \1eq{Pa5}
is added to the term 
$-\gamma_5 \sla{p_2}$ of \1eq{eq:SDEWTI-I}, and after use of the gap equation in \1eq{eq:GapEq}, one recovers precisely the term $-\gamma_5 S^{-1}(p_2)$ 
in \wtig. However, the second term, $V(p_1,p_2)$, 
does not admit an interpretation as  a genuine quark self-energy, leading to the obvious violation of the fundamental \wtig. 
\item Within the RL approximation, 
where $\g_\nu(q,r,-p)
\to \gamma_\nu$, the 
symmetry violating term $V(p_1,p_2)$
reduces to the required self-energy, 
since 
\begin{align}
V(p_1,p_2) \stackrel{\rm RL} {\longrightarrow} 
ig^2 C_f \int_q\gamma^\sigma S(q_1)\gamma_5\gamma^\nu \Delta_{\nu\sigma}(q)
= i \Sigma_{\rm {\s R \s L}}(p_1) \gamma_5  \,.
\label{theV}
\end{align}
Thus, diagram $(a_5^\mu)$ reproduces the \wtig in the RL approximation, but fails to do so
when the full kinematic structure of the 
quark-gluon vertex is maintained.  
\item One may recast the 
term $V(p_1,p_2)$ as the difference between the required quark self-energy, 
$i \Sigma(p_1)$, 
and a symmetry violating remainder, to be denoted by  
${\cal V}(p_1,p_2)$, 
through the addition and subtraction of the term 
$\g^\nu (q,p_1,-q_1)\gamma_5$,
\ie
\begin{align}
V(p_1,p_2) = &\, \underbrace{\left[-ig^2 C_f\int_q\gamma^\sigma S(q_1)\g^\nu(q,p_1,-q_1)\Delta_{\nu\sigma}(q)\right]}_{i\Sigma(p_1)}\gamma_5\\[1ex]
+ &\, \underbrace{ig^2 C_f\int_q\gamma^\sigma S(q_1)\left[\g^\nu(q,p_1,-q_1)\gamma_5+\gamma_5\g^\nu(q,p_2,-q_2)\right]\Delta_{\nu\sigma}(q)}_{\mathcal{V}(p_1,p_2):\,\textrm{symmetry violating remainder}}\,.
\label{thecalV}
\end{align}
Thus, one finally has that  
\begin{align}
P_\mu(a_5^\mu) = 
i\left[\Sigma(p_1)\gamma_5 + 
\gamma_5 \Sigma(p_2)\right] 
+ \mathcal{V}(p_1,p_2) \,,
\label{moreV}
\end{align}
with the symmetry violating 
remainder $\mathcal{V}(p_1,p_2)$ 
given in \1eq{thecalV}. Clearly, in the 
RL approximation, 
$\mathcal{V}(p_1,p_2) \sim
\gamma^{\nu} \gamma_5 +
\gamma_5 \gamma^{\nu} =0$. 
\end{enumerate}

It is now apparent that 
the gluon-axial-vector vertex $G^{\mu\nu}_5$, depicted in  \fig{fig:Elementary5}, 
must provide the 
{\it symmetry-restoring} contribution 
that will annihilate precisely 
the term $\mathcal{V}(p_1,p_2)$.
It turns out that 
this is indeed what happens, by virtue of the fundamental WTI satisfied by $G^{\mu\nu}_5$,
namely the \wtiG in \1eq{eq:G5WTI}. Indeed, using \wtiG and the expression for $(b_5^{\mu})$
given in \1eq{eq:b5mu}, we 
obtain directly 
\begin{align}\nonumber 
P_\mu(b_5^\mu) =&\, -ig^2C_f\int_q \gamma^\sigma S(q_1) \left[\g^\nu (q,p_1,-q_1)\gamma_5+\gamma_5 \g^\nu(q,p_2,-q_2)\right]
\Delta_{\nu\sigma}(q)\\[1ex]
=&\, - \mathcal{V}(p_1,p_2) \,.
\label{eq:symres}
\end{align}
Therefore, after the inclusion 
of the vertex $G^{\mu\nu}_5$, 
the key \wtig
is exactly fulfilled at the level of the SDE for $\ga^\mu$,  
\be
-P_\mu \left[\gamma_5\gamma^\mu+ (a^\mu_5) +(b^\mu_5)\right] 
= \is(p_1)\gamma_5+\gamma_5\is(p_2)\,. 
\ee

Note that the essence of this result 
was already stated in \cite{Chang:2009zb}, where the need of a vertex satisfying \wtiG in order to preserve the axial WTI at the level of the $\ga^\mu$ SDE was duly recognized.

%%%%%%%%%%%%%%%%%%%%%%%%%%%%%%%%%%%%%%%%%%%%%%%%%%%%
\section{Dynamical realization of the chiral limit}\label{sec:ChiralLim}

As explained in \sect{sec:WTI}, 
the pivotal relations in 
\2eqs{eq:chi1chi1}{eq:chi1chi3}
are a direct consequence of the \wtig. 
Given that, as shown in detail in \sect{sec:SDE}, the SDE that governs the vertex 
$\ga^\mu$ satisfies 
the \wtig {\it exactly},  
it is natural to expect that there will be a dynamical 
realization of \2eqs{eq:chi1chi1}{eq:chi1chi3}, 
recovered by taking the $P \to 0$ 
limit of 
diagrams $(a_5^{\mu})$ and $(b_5^{\mu})$ in \fig{fig:SDEPicture}. 
This particular demonstration is nontrivial, and furnishes additional valuable insights 
into this subject; 
the aim of this section is to show in detail how this important result emerges both within the RL approximation and in the general case.

%%%%%%%%%%%%%%%%%%%%%%%%%%%%%%%%
\subsection{Chiral limit in the RL approximation}

In order to fix the ideas, it is 
instructive to revert to the 
simpler case of the RL approximation, which involves 
only diagram $(a_5^{\mu})$, 
with $\g_\nu(q,r,-p)
\to \gamma_\nu$, \ie
\begin{align}
\ga^\mu =\gamma_5\gamma^\mu+ (a^\mu_5)_{\rm \s R \s L}\,,
\label{eq:SDEavRL}
\end{align}
with
\begin{align}
(a_5^\mu)_{\rm \s R \s L} =  -ig^2C_f\int_q \gamma^\sigma S(q_1)\ga^\mu (P,q_2,-q_1)S(q_2)\gamma^\nu \Delta_{\nu\sigma}(q) \,.
\label{arb}
\end{align}
When we contract both sides of 
\1eq{eq:SDEavRL} by $P_{\mu}$ 
and take the limit $P \to 0$ 
(with $p_1=p_2:=p$),  
the corresponding pole residues are isolated, while the tree-level term is annihilated. In particular,  
using 
\2eqs{eq:WTIP0-II}{chi0pp} on both sides, 
we get 
\begin{align}
[\chi_1(p^2)+\chi_3(p^2)\slashed{p}]\gamma_5 = 
-ig^2C_f
\int_q \gamma^\sigma S(q) 
\left[\chi_{1}(q^{2})+\chi_{3}(q^{2})\slashed{q}\right] \gamma_5
S(q)\gamma^\nu \Delta_{\nu\sigma}(q-p)\,,
\label{limprel}
\end{align}
where we have shifted the integration variable 
$q\to p+q$.

Using the basic property 
\mbox{$\gamma_5S(q)=S(-q)\gamma_5$}, and the elementary result 
\begin{align}
S(q)S(-q)=-c(q^2)\,,
\label{eq:SSminus}
\end{align}
with the function $c(q^2)$ defined in \1eq{eq:abc}, 
we find 
\begin{align}
\chi_1(p^2)+\chi_3(p^2)\slashed{p} = 
-ig^2C_f
\int_q \gamma^\sigma  
c(q^2) \left[\chi_{1}(q^{2})+\chi_{3}(q^{2})\slashed{q}\right] 
\gamma^\nu \Delta_{\nu\sigma}(q-p)  \,,
\label{parlc}
\end{align}
where the $\gamma_5$ 
has been canceled from both sides. 

It is now elementary to isolate two 
separate equations from \1eq{parlc}, 
by taking appropriate Dirac traces,
namely 
\begin{align}
\chi_1(p^2) = -\frac{ig^2 C_f}{4} \int_q 
c(q^2) \chi_{1}(q^{2}) 
\textrm{Tr}\left[\gamma^\sigma\gamma^\nu\right]
\Delta_{\nu\sigma}(q-p)
\,,
\label{chi1rl}
\end{align}
and  
\begin{align}
p^2\chi_3(p^2)
= \frac{ig^2 C_f}{4} \int_q 
c(q^2) \chi_{3}(q^{2}) 
\textrm{Tr}\left[\slashed{p}\gamma^\sigma \slashed{q}\gamma^\nu\right]
\Delta_{\nu\sigma}(q-p) 
\,.
\label{chi3rl}
\end{align}
It is clear at this point that \1eq{chi3rl} 
admits the trivial solution 
$\chi_3(p^2) =0$, in compliance with  \1eq{eq:chi1chi3}. 
In addition, substituting  
\1eq{eq:chi1chi1} 
into \1eq{chi1rl}, 
and using that 
\mbox{$b(q^2)=c(q^2)B(q^2)$}
[see \1eq{eq:abc}], one 
recovers precisely the 
dynamical equation governing $B(p^2)$ in the 
RL approximation, namely
the second relation in 
\1eq{eq:ABRL}. 
Thus, the dynamical equations 
(\ref{chi1rl}) and (\ref{chi3rl}),
derived from the SDE of the axial-vector vertex, are fully consistent with 
the symmetry relations 
(\ref{eq:chi1chi1})
and (\ref{eq:chi1chi3}).

%%%%%%%%%%%%%%%%%%%%%%%%%%%%%%%
\subsection{Chiral limit with full quark-gluon vertex}

Let us consider again 
the SDE of \1eq{eq:SDEav}
\begin{align}
\ga^\mu=\gamma_5\gamma^\mu+(a_5^\mu)+(b_5^\mu)\,,
\label{eq:SDEavII}
\end{align}
with $(a_5^\mu)$ and $(b_5^\mu)$  given by \2eqs{eq:a5mu}{eq:b5mu}, 
now maintaining inside $(a_5^{\mu})$ 
the full structure of 
the quark-gluon vertex
$\Gamma_{\nu}(q,r,-p)$, 
as given by 
\1eq{qgtens}.

Next, we contract both sides of 
\1eq{eq:SDEavII} by $P_{\mu}$, 
and take the limit $P \to 0$.
Evidently, the l.h.s. is exactly the same as that of 
\1eq{limprel}. On the r.h.s., 
the leading contributions of $(a_5^\mu)$ and $(b_5^\mu)$ are generated by the pole part of the vertices 
$\ga^\mu$ and $G_5^{\mu\nu}$,
respectively, 
\ie  
\begin{align}
\lim_{P\to 0}P_{\mu}(a_5^\mu) = &\, -ig^2 C_f \int_q \gamma^\sigma S(q)\left[\chi_1(q^2)+\chi_3(q^2)\slashed{q}\right]\gamma_5 S(q)\g^\nu(q',p,-q)\Delta_{\nu\sigma}(q')\,,\label{eq:a5muP0}\\[1ex]
\lim_{P\to 0}P_{\mu}(b_5^\mu) = &\, -2ig^2 C_f\int_q \gamma^\sigma S(q)\Gamma_2^\nu(q',p,-q)\Delta_{\nu\sigma}(q')\gamma_5\,,
\label{eq:b5muP0}
\end{align}
where we have used \1eq{eq:G5poleG2} to arrive at  \1eq{eq:b5muP0}, and we have set $q'=q-p$.

\begin{figure}[!t]
    \hspace*{-0.75cm}
    \includegraphics[scale=1]{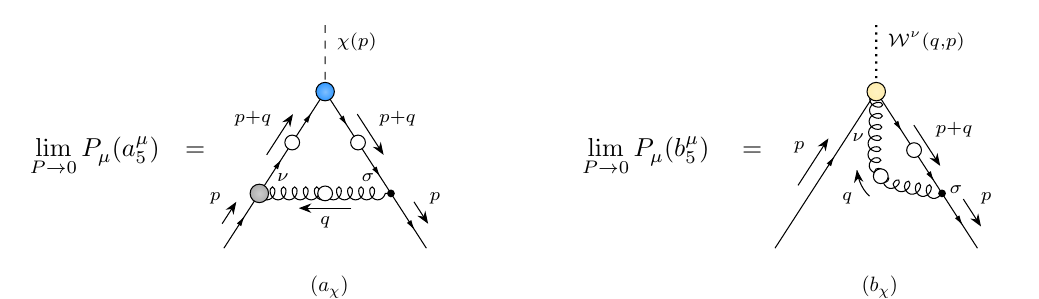}
    \caption{Diagrammatic representation of $(a_\chi)$ and $(b_\chi)$: the contributions 
    of diagrams ($a_5^\mu)$ and $(b_5^\mu)$, respectively,  
    in the limit $P\to 0$.}
    \label{fig:abchi}
\end{figure}

Using \1eq{eq:SSminus}, 
the decomposition 
of the quark-gluon vertex in 
\1eq{eq:QgOddEven}, and 
the relation in \1eq{eq:QgOddEvengamma5}, 
we may bring 
the $\gamma_5$ of \1eq{eq:a5muP0}
all the way to the right, namely 
\begin{align}
\lim_{P\to 0}P_{\mu}(a_5^\mu) = ig^2 C_f \int_q \gamma^\sigma c(q^2)\left[\chi_1(q^2)+\chi_3(q^2)\slashed{q}\right] \left[\ge^\nu(q',p,-q)-\go^\nu(q',p,-q)\right]\Delta_{\nu\sigma}(q')\gamma_5\,.
\end{align}
Therefore, after canceling the $\gamma_5$ from both sides, we arrive at the result 
\be\label{eq:dynchiII}
\chi_1(p^2)+\chi_3(p^2)\slashed{p} = (a_\chi)+(b_\chi)\,,
\ee
with
\begin{align}
(a_\chi) =&
\,ig^2 C_f\int_q \gamma^\sigma c(q^2)\left[\chi_1(q^2)+\chi_3(q^2)\slashed{q}\right]\left[\ge^\nu(q',p,-q)-\go^\nu(q',p,-q)\right]\Delta_{\nu\sigma}(q')\,,\label{eq:achi}\\[1ex]
(b_\chi) = &\, -2ig^2 C_f\int_q \gamma^\sigma S(q)
\Gamma_2^\nu(q',p,-q)
\Delta_{\nu\sigma}(q')\label{eq:bchi}\,,
\end{align}
which may be diagrammatically interpreted as shown in \fig{fig:abchi}.

The two individual equations for 
$\chi_1(p^2)$ and $\chi_3(p^2)$
are given by 
\begin{align}
4\chi_1(p^2) =&\, \textrm{Tr}(a_\chi)+\textrm{Tr}(b_\chi) \,,
\label{eq:systema}
\\[1ex]
4p^2\chi_3(p^2) =&\,  \textrm{Tr}\left[\slashed{p}(a_\chi)\right]+\textrm{Tr}\left[\slashed{p}(b_\chi)\right] \,,
\label{eq:systemb}
\end{align}
It is relatively straightforward to determine the four traces 
appearing in \2eqs{eq:systema}{eq:systemb},
namely 
\begin{align}\nonumber 
\textrm{Tr}(a_\chi) = &\, ig^2C_f\int_q c(q^2)\chi_3(q^2)\textrm{Tr}\left[\gamma^\sigma\slashed{q}\ge^\nu(q',p,-q)\right]\Delta_{\nu\sigma}(q')\\[1ex]
-&\, ig^2C_f\int_q c(q^2)\chi_1(q^2)\textrm{Tr}\left[\gamma^\sigma\go^\nu(q',p,-q)\right]\Delta_{\nu\sigma}(q')\,,
\label{tra}
\\[2ex]\nonumber 
\textrm{Tr}(b_\chi) =&\, 
-2ig^2 C_f\int_q\textrm{Tr}\left[\gamma^\sigma S(q)\ge^\nu(q',p,-q)\right]\Delta_{\nu\sigma}(q')\\[1ex]
=&\,-2ig^2 C_f\int_q a(q^2)\textrm{Tr}\left[\gamma^\sigma\slashed{q}\ge^\nu(q',p,-q)\right]\Delta_{\nu\sigma}(q')\,,
\label{trb}
\end{align}
and 
\begin{align}\nonumber 
\textrm{Tr}\left[\slashed{p}(a_\chi)\right]  = &\, ig^2 C_f\int_q c(q^2)\chi_1(q^2)\textrm{Tr}\left[\slashed{p}\gamma^\sigma\ge^\nu(q',p,-q)\right]\Delta_{\nu\sigma}(q')\\[1ex]
 - &\, ig^2 C_f\int_q c(q^2)\chi_3(q^2)\textrm{Tr}\left[\slashed{p}\gamma^\sigma\slashed{q}\go^\nu(q',p,-q)\right]\Delta_{\nu\sigma}(q')\,,
\label{trpa}\\[2ex] \nonumber 
\textrm{Tr}\left[\slashed{p}(b_\chi)\right]  = &\, -2ig^2 C_f\int_q\textrm{Tr}\left[\slashed{p}\gamma^\sigma S(q)\ge^\nu(q',p,-q)\right]\Delta_{\nu\sigma}(q')\\[1ex]
= &\, -2ig^2 C_f\int_q b(q^2)\textrm{Tr}\left[\slashed{p}\gamma^\sigma \ge^\nu(q',p,-q)\right]\Delta_{\nu\sigma}(q')\,.
\label{trpb}
\end{align}

Note that in computing 
$\textrm{Tr}(b_\chi)$
and
$\textrm{Tr}\left[\slashed{p}(b_\chi)\right]$
we have 
used the parametrization introduced in \1eq{eq:S}, setting 
\mbox{$S(q) = a(q^2)\slashed{q}+b(q^2)$}.

We are now in a position to verify that the system in \2eqs{eq:systema}{eq:systemb} is compatible with 
the key symmetry relations 
(\ref{eq:chi1chi1})
and (\ref{eq:chi1chi3}), namely 
\mbox{$\{\chi_1(p^2), \chi_3(p^2)\} =
\{2B(p^2), 0\}$}. 

To that end, we first focus on \1eq{eq:systemb},
and note that the substitution 
$\chi_1(q^2) = 2B(q^2)$ into 
the first term of  
$\textrm{Tr}\left[\slashed{p}(a_\chi)\right]$ in \1eq{trpa}
triggers a crucial cancellation.
Specifically, using \1eq{eq:abc},
we have that 
\mbox{$c(q^2)\chi_1(q^2) = 2 c(q^2) B(q^2) = 2b(q^2)$},
and so 
$\textrm{Tr}\left[\slashed{p}(b_\chi)\right]$ cancels exactly 
against this term, namely 
\be
\textrm{Tr}\left[\slashed{p}(a_\chi)\right] + \textrm{Tr}\left[\slashed{p}(b_\chi)\right] = 
ig^2C_f\int_q c(q^2)\chi_3(q^2)\textrm{Tr}\left[\gamma^\sigma\slashed{q}\ge^\nu(q',p,-q)\right]\Delta_{\nu\sigma}(q') \,.
\label{trpab}
\ee
Thus, by virtue of \1eq{trpab}, 
\1eq{eq:systemb} becomes 
\be
4p^2\chi_3(p^2) =
-ig^2C_f\int_q c(q^2)\chi_3(q^2)\textrm{Tr}\left[\slashed{p}\gamma^\sigma\slashed{q}\go^\nu(q',p,-q)\right]\Delta_{\nu\sigma}(q') \,,
\ee
which obviously admits the 
trivial solution $\chi_3(p^2) =0$,
as dictated by the symmetry.
Note that the inclusion of graph 
$(b_5^{\mu})$ is instrumental for 
achieving this result;  
indeed, the 
contribution of 
diagram $(a_5^{\mu})$ alone 
fails to vanish when 
$\chi_3 =0$, because of the term 
$\int_q c(q^2)\chi_1(q^2)\textrm{Tr}\left[\slashed{p}\gamma^\sigma\ge^\nu(q',p,-q)\right]\Delta_{\nu\sigma}(q')$, which 
vanishes in the RL 
(because $\Gamma_2^\nu =0$,  trivially), but is nonvanishing 
in general.  

Turning to \1eq{eq:systema}, 
and substituting 
\mbox{$\{\chi_1(p^2), \chi_3(p^2)\} =
\{2B(p^2), 0\}$} into 
\2eqs{tra}{trb}, we arrive at   
\bea\label{eq:chi1}
     B(p^2) & = & -\frac{ig^2C_f}
     {4}\int_q b(q^2)\textrm{Tr}\left[\gamma^\sigma\go^{\nu}(q',p,-q)\right]\Delta_{\nu\sigma}(q')\nonumber\\
     \nonumber\\
     && -\frac{ig^2C_f}{4}\int_q a(q^2)\textrm{Tr}\left[\gamma^\sigma\slashed{q}\ge^{\nu}(q',p,-q)\right]\Delta_{\nu\sigma}(q')\,,
\eea
which is precisely \1eq{eq:BEq}, the dynamical equation for $B(p^2)$ obtained from the gap equation. 

The results of this section 
demonstrate clearly 
that the two key equations 
controlling the axial-vector vertex, 
namely its SDE and its WTI,
are completely compatible 
with each other.

%%%%%%%%%%%%%%%%%%%%%%%%%%%%%%%%%%%%%%%%%%%%%%%%%%%%
\section{Unfolding the \texorpdfstring{$G_5^{\mu\nu}$}{gav}: One-loop dressed approximation}\label{sec:OpenG5}

Up until this point, 
the SDE 
for the gluon-axial-vector vertex $G_5^{\mu\nu}$
has been treated exactly, without resorting to any type of approximation.
However, for practical
purposes, we need either a diagrammatic expansion for $G_5^{\mu\nu}$, or an Ansatz that 
capitalizes on the \wtiG 
that this vertex satisfies, see discussion at the end of this section.

\begin{figure}[t!]
    \centering
    \includegraphics[scale=1]{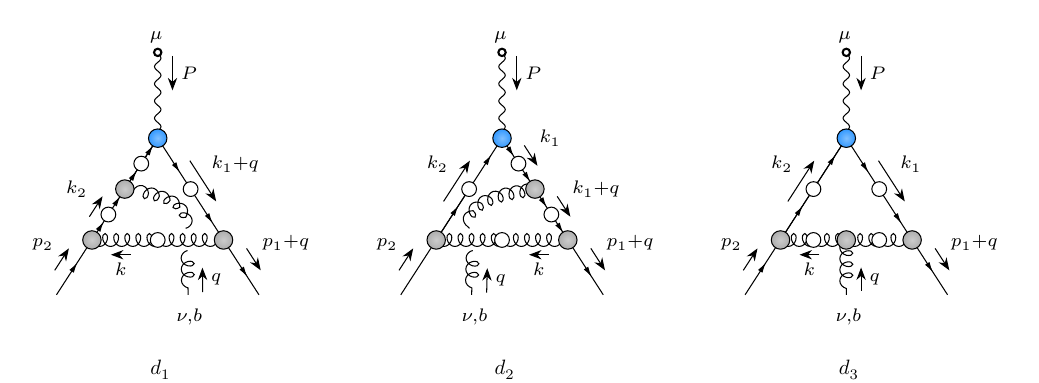}
    \caption{One-loop dressed diagrams contributing to the gluon-axial-vector vertex SDE, in the lowest-order of its dressed-loop expansion, and not involving $G_5^{\mu\nu}$ itself.}
    \label{fig:G5diagrams}
\end{figure}

\begin{figure}[t!]
    \centering
    \includegraphics[scale=1]{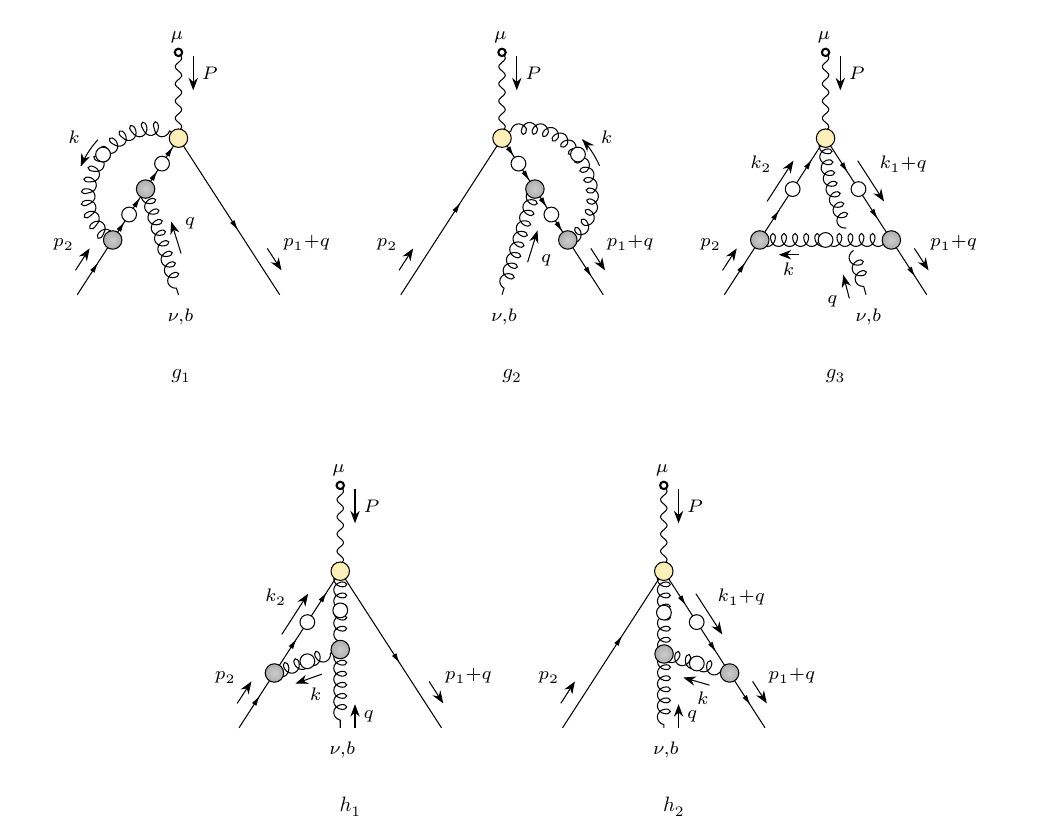}
    \caption{One-loop dressed diagrams contributing to the gluon-axial-vector vertex SDE in the lowest-order of its dressed-loop expansion, and involving $G_5^{\mu\nu}$ itself.}
    \label{fig:G5selfdiagrams}
\end{figure}

Opting for the 
former possibility, 
in what follows we  
consider 
the {\it ``one-loop dressed''} approximation
to $G_5^{\mu\nu}$, 
consisting of 
one-loop diagrams   
that 
contain fully-dressed propagators and vertices, as shown in Figs.~\ref{fig:G5diagrams} and \ref{fig:G5selfdiagrams}. 
Note that this type of 
truncation has 
already been discussed in 
\sect{sec:GapEq}, in the context of the SDE for the quark-gluon vertex. 
Therefore, in complete analogy to the $\g_\mu^{({\bf \s 1})}$ introduced in \1eq{Goneup}, 
the gluon-axial-vector vertex 
emerging from this approximation 
will be denoted by $\GNEW$.

The vertex $\GNEW$ 
can be expressed as 
\be\label{eq:G5SDE}
\GNEW=\mathcal{A}_{5\mu\nu}+\mathcal{B}_{5\mu\nu}\,,
\ee
where
\be
\mathcal{A}_5^{\mu\nu}:= d_1^{\mu\nu}+d_2^{\mu\nu}+ g_1^{\mu\nu}
+ g_2^{\mu\nu}
+ g_3^{\mu\nu}
\,,\qquad \mathcal{B}_5^{\mu\nu}:= d_3^{\mu\nu} +
h_1^{\mu\nu} + h_2^{\mu\nu}\,,
\label{eq:AandB}
\ee
and $d_i^{\mu\nu}$, $g_i^{\mu\nu}$ and $h_i^{\mu\nu}$ label the
contributions shown in \fig{fig:G5diagrams} and \fig{fig:G5selfdiagrams}. The obvious 
distinction between these two sets is that 
the diagrams belonging to $\mathcal{A}_5^{\mu\nu}$ do not contain a three-gluon vertex, while those of   
$\mathcal{B}_5^{\mu\nu}$ do;  we will therefore refer to these two sets as ``abelian'' and ``non-abelian'', respectively. 
In addition, 
note that 
the susbet of diagrams shown in \fig{fig:G5selfdiagrams} 
contain the vertex 
$G_5^{\mu\nu}$ itself as their basic ingredient.

We now turn to the key question of  
what happens to  
the \wtiG 
when $G_{5\mu\nu}$ is 
approximated by 
\2eqs{eq:G5SDE}{eq:AandB}, 
\ie 
$G_{5\mu\nu}\to \GNEW$.
In order to address this issue, 
we will contract by $P_{\mu}$
all diagrams given by 
\fig{fig:G5diagrams} and \fig{fig:G5selfdiagrams}.
Quite interestingly,
the answer that will emerge 
is {\it precisely}
the r.h.s. of \wtiG, 
but with  
$\g_{\nu} \to 
\g^{({\bf \s 1})}_{\nu}$.
In other words, 
the one-loop dressed versions of 
$G_{5\mu\nu}$ and $\g_{\nu}$
are connected by the correct WTI.

We start by considering the set of abelian diagrams 
belonging to the set 
$\mathcal{A}_5^{\mu\nu}$. Contracting by $P_\mu$ and using \wtig, we write
\be\label{eq:Pd12}
P_\mu d_1^{\mu\nu}=-L_1^\nu-L_3^\nu\,,\qquad P_\mu d_2^{\mu\nu}=-L_2^{\nu}-L_4^{\nu}\,,
\ee
where
\bea\label{eq:Z1234}
      L_1^\nu & = & g^2\kappa_a \int_k \g^\beta(-k,k'_1,-q_1)\gamma_5 S(k'_2)\g^\nu(q,k_2,-k'_2)S(k_2)\g^\alpha(k,p_2,-r_2)\Delta_{\alpha\beta}(k)\,,\nonumber\\
     \nonumber\\
     L_2^\nu & = &  g^2\kappa_a\int_k \g^\beta(-k,k'_1,-q_1)S(k'_1)\g^\nu(q,k_1,-k'_1)S(k_1)\gamma_5 \g^\alpha(k,p_2,-k_2)\Delta_{\alpha\beta}(k)\,,\nonumber\\
    \nonumber\\
     L_3^\nu & = & g^2\kappa_a\int_k \g^\beta(-k,k'_1,-q_1)S(k'_1)\gamma_5 \g^\nu(q,k_2,-k'_2)S(k_2)\g^\alpha(k,p_2,-k_2)\Delta_{\alpha\beta}(k)\,,\nonumber\\
     \nonumber\\
     L_4^\nu & = & g^2\kappa_a \int_k \g^\beta(-k,k'_1,-q_1)S(k'_1)\g^\nu(q,k_1,-k'_1)\gamma_5 S(k_2)\g^\alpha(k,p_2,-k_2)\Delta_{\alpha\beta}(k)\,,
\eea
with $k'_i:=k_i+q$ and $\kappa_a:=C_f-C_A/2$. 

A similar calculation 
may be carried out 
for the diagrams $\lbrace g\rbrace_i$, \ie
\bea\label{eq:Pg123}
P_\mu g_1^{\mu\nu} & = & L^\nu_1+g^2\kappa_a\int_k\gamma_5\g^\beta(-k,k'_2,-q_2)S(k'_2)\g^\nu(q,k_2,-k'_2) S(k_2)\g^\alpha(k,p_2,-k_2)\Delta_{\alpha\beta}(k)\,,\nonumber\\
\nonumber\\
P_\mu g_2^{\mu\nu} & = & L^\nu_2+\displaystyle g^2\kappa_a\int_k\g^\beta(-k,k'_1,-q_1) S(k'_1)\g^\nu(q,k_1,-k'_1) S(k_1)\g^\alpha(k,p_1,-k_1)\gamma_5\Delta_{\alpha\beta}(k)\,,\nonumber\\
\nonumber\\
P_\mu g_3^{\mu\nu} & = & L^\nu_3+L^\nu_4\,.
\eea

When these five contributions
are added up,
a number of key cancellations takes place, leaving us with two terms, namely
\bea\label{eq:Pd12g123}
     P_\mu \mathcal{A}_5^{\mu\nu} & = & \underbrace{g^2\kappa_a\int_k\g^\beta(-k,k'_1,-q_1)S(k'_1)\g^\nu(q,k_1,-k'_1)S(k_1)\g^\alpha(k,p_1,-k_1)\Delta_{\alpha\beta}(k)}_{(c_1^\nu){\rm\,of\,\fig{fig:QGSDE}\, at\, }\lbrace r,-p\rbrace=\lbrace p_1,-q_1\rbrace }\gamma_5\nonumber\\
     \nonumber\\
     & + & \gamma_5 \underbrace{g^2\kappa_a\int_k \g^\beta(-k,k'_2,-q_2)S(k'_2)\g^\nu(q,k_2,-k'_2)S(k_2)\g^\alpha(k,p_2,-k_2)\Delta_{\alpha\beta}(k)}_{(c_1^\nu){\rm\,of\, \fig{fig:QGSDE}\, at\, }\lbrace r,-p\rbrace=\lbrace p_2,-q_2\rbrace }\,,\nonumber\\
\eea
or, diagrammatically,
\be\label{eq:Pd12g123graph}
-iP_\mu \mathcal{A}_5^{\mu\nu}=\left(\raisebox{-1cm}{\includegraphics[scale=1]{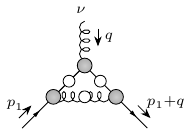}}\right)\gamma_5\quad +\quad\gamma_5\left(\raisebox{-1cm}{\includegraphics[scale=1]{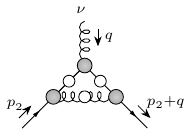}}\right)~.
\ee
Next, we turn to the non-abelian subset $\mathcal{B}^{\mu\nu}_5$, and follow 
completely analogous steps. 
In particular, 
\be\label{eq:Pd3}
P_\mu d_3^{\mu\nu}=-L_5^\nu-L_6^\nu\,,
\ee
where
\bea\label{eq:Y12}
L_5^\nu & = & -ig^2\kappa_b\int_k\g^\beta(-k',k_1,-q_1)\gamma_5 S(k_2)\g^\alpha(k,p_2,-k_2)\g^{\nu\rho\sigma}(q,-k,k')\Delta_{\alpha\rho}(k)\Delta_{\sigma\beta}(k')\,,\nonumber\\
\nonumber\\
\displaystyle L_6^\nu & = & -ig^2\kappa_b\int_k\g^\beta(-k',k_1,-q_1)S(k_1)\gamma_5\g^\alpha(k,p_2,-k_2)\g^{\nu\rho\sigma}(q,-k,k')\Delta_{\alpha\rho}(k)\Delta_{\sigma\beta}(k')\,,\nonumber\\
\eea
with $k':=k-q$ and $\kappa_b :=iC_A/2$, and 
\bea\label{eq:Ph12}
\displaystyle P_\mu h_1^{\mu\nu} & = & L^\nu_5-ig^2\kappa_b\int_k \gamma_5 \g^\beta(-k',k_2,-q_2) S(k_2)\g^\alpha(k,p_2,-k_2)\g^{\nu\rho\sigma}(q,-k,k')\Delta_{\alpha\rho}(k)\Delta_{\sigma\beta}(k')\,,\nonumber\\
\nonumber\\
P_\mu h_2^{\mu\nu} & = & L_6^\nu-ig^2\kappa_b\int_k \g^\beta(-k',k_1,-q_1)S(k_1)\g^\alpha(k,p_1,-k_1)\gamma_5 \g^{\nu\rho\sigma}(q,-k,k')\Delta_{\alpha\rho}(k)\Delta_{\sigma\beta}(k')\,.\nonumber\\
\eea
As in the abelian case, an extensive 
cancellation takes place when the 
terms in \2eqs{eq:Y12}{eq:Ph12} are summed up: 
the entire contribution given by $d_3^{\mu\nu}$ is exactly canceled by parts of $h_1^{\mu\nu}$ and $h_2^{\mu\nu}$, leaving us with 
\begin{align} \nonumber 
-P_\mu \mathcal{B}_5^{\mu\nu} = &\, \underbrace{ig^2\kappa_b\int_k \g^\beta(-k',k_1,-q_1)S(k_1)\g^\alpha(k,p_1,-k_1)\g^{\nu\rho\sigma}(q,-k,k')\Delta_{\alpha\rho}(k)\Delta_{\sigma\beta}(k')}_{(c_2^\nu){\rm\,of\,\fig{fig:QGSDE}\, at\, }\lbrace r,-p\rbrace=\lbrace p_1,-q_1\rbrace }\gamma_5\\[4ex]
&\,+ \gamma_5\underbrace{ig^2\kappa_b\int_k\g^\beta(-k',k_2,-q_2)S(k_2)\g^\alpha(k,p_2,-k_2)\g^{\nu\rho\sigma}(q,-k,k')\Delta_{\alpha\rho}(k)\Delta_{\sigma\beta}(k')}_{(c_2^\nu){\rm\,of\,\fig{fig:QGSDE}\, at\, }\lbrace r,-p\rbrace=\lbrace p_2,-q_2\rbrace }\,,
\label{eq:Pd3h12}
\end{align}
or, diagrammatically,
\begin{align}
-iP_\mu\mathcal{B}_5^{\mu\nu}=\left(\raisebox{-1cm}{\includegraphics[scale=1]{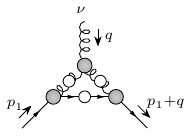}}\right)\gamma_5\quad +\quad \gamma_5\left(\raisebox{-1cm}{\includegraphics[scale=1]{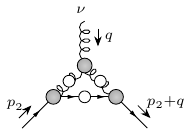}}\right)\,.
\label{eq:Pd3h12graph}
\end{align}
Evidently, now 
the non-abelian part of the exact \wtiG is formed, where
the quark-gluon vertices are   
given by the non-abelian diagram of the 
SDE in \fig{fig:QGSDE}.

Thus, combining  \2eqs{eq:Pd12g123graph}{eq:Pd3h12graph}, we arrive at 
\begin{align}\nonumber 
     -iP^\mu \GNEW(P,q,p_2,-q_1) = & \,\enskip\enskip\left(\raisebox{-1cm}{\includegraphics[scale=1]{abp1.pdf}}+\raisebox{-1cm}{\includegraphics[scale=1]{nabp1.pdf}}\right)\gamma_5\\[2ex]
     &\,+ \gamma_5\left(\raisebox{-1cm}{\includegraphics[scale=1]{abp2.pdf}}+\raisebox{-1cm}{\includegraphics[scale=1]{nabp2.pdf}}\right)\,.
\label{centres}
\end{align}
The result in \1eq{centres}
is rather striking: $\GNEW$, the 
one-loop dressed approximation of $G_{5\mu\nu}$,
satisfies the $[{\rm WTI}]_{G_5}$ 
of \1eq{eq:G5WTInotree}, 
provided that the quark-gluon vertex is determined from the one-loop dressed version 
of its own SDE, shown in \fig{fig:QGSDE}, to wit
\begin{align}
-iP^\mu \GNEW(P,q,p_2,-q_1)=\g_{\!\s Q\,\nu}^{({\bf \s 1})}(q,p_1,-q_1)\gamma_5+\gamma_5\g_{\!\s Q\, \nu}^{({\bf \s 1})}(q,p_2,-q_2)\,.
\label{eq:G5WTIcheck}
\end{align}
Clearly, as already mentioned in \sect{sec:WTI},  the tree-level contribution to the quark-gluon vertex may be added to the 
r.h.s. of \1eq{eq:G5WTIcheck}
``for free'', leading to 
\begin{align}
-iP^\mu \GNEW(P,q,p_2,-q_1)= \g^{({\bf \s 1})}_{\nu}(q,p_1,-q_1)\gamma_5+\gamma_5\g^{({\bf \s 1})}_{\nu}(q,p_2,-q_2)\,,
\label{eq:G5WTIcheckII}
\end{align}
which is precisely the 
one-loop dressed realization of the \wtiG.

We end this section by pointing out that, 
given the pivotal function  
of the vertex $G_5^{\mu\nu}$, 
it is important to consider 
alternative procedures for 
its determination.
In particular, since 
the \wtiG  
is a most prominent feature of this vertex,
one may opt for a  
gauge-technique type 
of construction~\cite{Salam:1963sa,Salam:1964zk,Delbourgo:1977jc,Delbourgo:1977hq,Abbott:1979dt,Ball:1980ay}.
This general method has been 
employed in numerous occasions 
in the literature, for a variety of vertices, 
see \eg\cite{Binosi:2011wi,
Aguilar:2014lha, 
Aguilar:2016lbe}. 
Through this well-defined 
procedure,  
the non-transverse part of 
$G_5^{\mu\nu}$ may be reconstructed exactly from the  
\wtiG, while the 
transverse part remains 
undetermined. 

%%%%%%%%%%%%%%%%%%%%%%%%%%%%%%%%%%%%%%%%%%%%%%%%%%%%
\section{Symmetry-preserving  truncation of  the vertex  
\texorpdfstring{$\ga^\mu$}{g5}}\label{sec:sdetrun}

It is clear that 
the one-loop dressed approximation
implemented at the level of 
$G_5^{\mu\nu}$ must be self-consistently incorporated
into the 
SDE satisfied by 
$\ga^\mu$, namely  
\1eq{eq:SDEav}, together with 
\2eqs{eq:a5mu}{eq:b5mu}. 
In particular, the 
effective replacement 
$G_{5\mu\nu}\to \GNEW$ 
imposed at the level of graph ($b_5^{\mu}$) must be 
accompanied by analogous adjustments, in order for the 
fundamental $[{\rm WTI}]_{G_5}$ 
to be maintained intact. 

In order to elucidate this point,
we remind the reader that, 
as shown 
in \sect{sec:SDE}, 
the contribution of $G_5^{\mu\nu}$
inside ($b_5^{\mu}$)
is instrumental for canceling the 
symmetry violating contribution
originating from ($a_5^{\mu}$).
This latter contribution, 
denoted by $\mathcal{V}(p_1,p_2)$,
contains precisely the r.h.s. of the $[{\rm WTI}]_{G_5}$, 
see \1eq{thecalV}. 
Evidently, if the effective substitution 
$G_{5\mu\nu}\to \GNEW$ 
is carried out in graph ($b_5^{\mu}$), 
then the one-loop version of 
the $[{\rm WTI}]_{G_5}$ in 
\1eq{eq:G5WTIcheckII}
is triggered. 
Therefore, in order for the 
aforementioned crucial cancellation between the two graphs 
to still go through, 
the corresponding replacement 
$\g_{\nu} \to 
\g^{({\bf \s 1})}_{\nu}$
must be carried out in 
graph ($a_5^{\mu}$). 

Thus, one arrives at the following truncated version of the SDE
for $\ga^\mu$, 
\begin{align}
\ga^\mu =\gamma_5\gamma^\mu+ (a^\mu_5)+(b^\mu_5)\,,
\label{eq:SDEav-trunc}
\end{align}
where
\begin{align}
     (a_5^\mu)  =&\, -ig^2C_f\int_q \gamma^\sigma S(q_1)\ga^\mu (P,q_2,-q_1)S(q_2) \g^{({\bf \s 1})}_{\nu}(q,p_2,-q_2)\Delta^{\nu}_{\sigma}(q)~,\label{eq:a5mu-trunc}
     \\[1ex]
     (b_5^\mu)  = &\, -g^2C_f\int_q \gamma^\sigma S(q_1)
     G_{5\nu}^{({\bf \s 1})\mu}(P,q,p_2,-q_1)\Delta^{\nu}_{\sigma}(q)\,.
\label{eq:b5mu-trunc}
\end{align}
It is important to emphasize that, while graph $(a_5^\mu)$ in \1eq{eq:a5mu-trunc} is ``one-loop dressed'', $(b_5^\mu)$ is clearly 
``two-loop dressed'', because the one-loop dressed $\GNEW$ is nested inside an additional integration over the gluon momentum
$q$ entering in it. 
In that sense, there is no 
self-consistent one-loop dressed treatment of the vertex 
$\Gamma_5^{\mu}$, except within the RL approximation;
within the confines of the present framework, 
any meaningful attempt to go beyond RL requires a beyond one-loop dressed treatment. 

Let us now turn to the 
$[{\rm WTI}]_{\Gamma_{\!5}}$ satisfied by the truncated vertex
$\ga^\mu$, defined through 
\3eqs{eq:SDEav-trunc}{eq:a5mu-trunc}
{eq:b5mu-trunc}. 
It is clear from the above analysis 
that the \wtig is satisfied, provided that we carry out the substitution 
$\g_{\mu}(q-p,p,-q) \to 
\g^{({\bf \s 1})}_{\mu}(q-p,p,-q)$ inside 
\1eq{eq:SeflEnergy}, namely 
the 
equation defining the quark self-energy
$\Sigma(p)$.
Following steps identical to those of \sect{sec:GapEq}, we can derive the dynamical equations for the quark dressing functions $A(p^2)$ and $B(p^2)$. Those are exactly as in  \2eqs{eq:AEq}{eq:BEq}, except for the substitution $\go\to \go^{({\bf \s 1})}$ and $\ge\to \ge^{({\bf \s 1})}$, defined by 
\begin{align}
\g_{\!1\nu}^{({\bf \s 1})}(q,p,-p-q) =\gamma_\nu+\left[\raisebox{-1cm}{\includegraphics[scale=1]{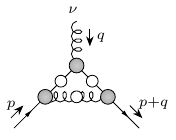}}+\raisebox{-1cm}{\includegraphics[scale=1]{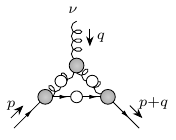}}\right]_{\textrm{odd}}\,,
\label{eq:gamma11}
\end{align}
and
\begin{align}
\g_{\!2\nu}^{({\bf \s 1})}(q,p,-p-q) =\left[\raisebox{-1cm}{\includegraphics[scale=1]{a0.pdf}}+\raisebox{-1cm}{\includegraphics[scale=1]{na0.pdf}}\right]_{\textrm{even}}\,.
\label{eq:gamma12}
\end{align}
%

%%%%%%%%%%%%%%%%%%%%%%%%%%%%%%%%%%
\section{Discussion and conclusions}\label{sec:Disc}

In the present work we have developed a novel approach
that aims at a self-consistent treatment of the functional equations associated with the physics of mesons in the non-singlet sector.

The cornerstone of this formalism is the exact SDE of the axial-vector vertex $\ga^{\mu}$, formulated 
from the viewpoint of the antiquark leg.
This alternative 
re-organization 
of the quantum corrections 
is described in terms 
of a new vertex, 
denominated gluon-axial-vector vertex, 
and denoted by 
$G_5^{\mu\nu}$. 
This vertex is nested inside 
the SDE of $\ga^{\mu}$, and  
satisfies its own 
WTI, which relates it 
to the quark-gluon vertex. 
As a result, the 
WTI of the axial-vector vertex is enforced directly at the level of 
the SDE. In particular, 
the gap equation governing 
the quark propagator appearing on the r.h.s. of this WTI involves the full 
quark-gluon vertex. 

Information on the 
BSE of the pion is obtained by taking 
the limit $P \to 0 $ of the above SDE, 
where $P$ is the momentum of the axial-vector current. 
In particular, as it is well-known, when the chiral symmetry is dynamically 
broken, the axial-vector vertex possesses a 
(longitudinally coupled)
pole in $P^2$, whose 
residue is identified with the BSA of the pion. 
By virtue of 
the aforementioned results, 
the known relation between the quark mass function, 
obtained from the gap equation with a full $\Gamma_{\mu}$, and the amplitude of the pion,
is precisely fulfilled. 

The developed framework 
offers a rather tight control 
on the interlocked equations describing 
symmetry and dynamics. As a result, it facilitates the 
design and 
implementation of simpler calculational methods.  
In particular, 
a concrete  
truncation scheme 
for the SDE-BSE system has 
been identified in 
\sect{sec:OpenG5}, referred 
to as ``one-loop dressed'' 
approximation; it incorporates a completely nontrivial quark-gluon vertex $\Gamma_{\mu}$, which is 
fully compatible with chiral symmetry, as expressed through the axial WTI. 

It is worth pointing out 
that, throughout all algebraic manipulations of our analysis, 
the gluon propagator 
$\Delta_{\mu\nu}$ 
remained completely 
inert. Therefore, 
the formal aspects of this 
work persist  
for any value of the 
gauge-fixing parameter 
$\xi$, or even within gauge-fixing schemes other than the covariant gauges. 
Even though, for practical purposes, the Landau gauge 
is undoubtedly the standard 
choice, the validity of our 
considerations for arbitrary gauges is a welcome feature. 

It is important to emphasize that our approach has been developed strictly in the 
limit of vanishing current quark masses. Its 
extension to arbitrary  
current masses is expected to proceed 
in a similar fashion, once 
the modifications to
the key WTIs have been appropriately taken into account. 
In particular, 
in the presence of current masses, 
the WTIs contain
additional terms involving the axial vertex 
$\Gamma_5$, see, \eg Eq.\nobreak\thinspace(11.217) in~ \cite{Itzykson:1980rh}, or 
Eq.\nobreak\thinspace(14) in~\cite{Maris:1997hd}.
The vertex $\Gamma_5$ is
associated with the 
axial current $j_5=i\bar{\psi}\gamma_5\psi$, which 
satisfies the relation $\partial_\mu j_5^{\mu,a}=-2mt^a j_5$.
We hope to present the 
completion of this task 
in the near future. 

The practical implementation 
of the one-loop-dressed 
approximation mentioned above 
may not be easy to carry out, 
due to the presence of the 
vertex $G_5^{\mu\nu}$ inside some of the relevant diagrams, see \fig{fig:G5selfdiagrams}. 
It turns out, however, that 
a further simplification is possible: one may omit the aforementioned diagrams,  
keeping only those  
in \fig{fig:G5diagrams}, 
provided that  the SDE of the quark-gluon vertex, \fig{fig:QGSDE}, is 
treated in the so-called ``symmetric" limit.  
This treatment amounts to 
replacing the full 
momentum-dependence of the 
form factors  
in the quark-gluon SDE 
by the choice $q^2=r^2 =p^2$, 
see, \eg\cite{Gao:2021wun}.
This additional simplification  is expected to lead to a more tractable version of the pion BSE, apt for numerical exploration. We expect to report progress 
in this direction in a forthcoming communication. 

%%%%%%%%%%%%%%%%%%%%%%%%%%%%%%%%%%%
\section*{Acknowledgments}
\label{sec:Acknow}

The work of A.S.M., J.M.M.C. and J.P. is funded by the Spanish MICINN grants PID2020-113334GB-I00 and PID2023-151418NB-I00, the Generalitat Valenciana grant CIPROM/2022/66, and CEX2023-001292-S by MCIU/AEI. J.P. is supported in part by the EMMI visiting grant of the ExtreMe Matter Institute EMMI at the GSI, Darmstadt, Germany. J.M.P. is funded by the Deutsche Forschungsgemeinschaft (DFG, German Research Foundation) under Germany’s Excellence Strategy EXC 2181/1 - 390900948 (the Heidelberg STRUCTURES Excellence Cluster) and the Collaborative Research Centre SFB 1225 - 273811115 (ISOQUANT).

%%%%%%%%%%%%%%%%%%%%%%%%%%%%%%%%%%%
\appendix
\section{Derivation of the WTIs}\label{app:WTIproof}

In this Appendix we derive the 
key WTIs satisfied by the 
axial-vector vertex, $\ga^\mu$, 
and the 
gluon-axial-vector vertex, 
$G_5^{\mu\nu}$,
namely the $[{\rm WTI}]_{\Gamma_{\!5}}$ of \1eq{eq:WTI5} 
and the $[{\rm WTI}]_{G_5}$ of 
\1eq{eq:G5WTI}, 
respectively. 

%%%%%%%%%%%%%%%%%%%%%%%%%%%%%%%%%%
\subsection{The \texorpdfstring{$[{\rm WTI}]_{\Gamma_{\!5}}$}{wtig5}}

The starting point 
for the derivation of the 
required WTIs is 
the master formula 
given in Eq.\nobreak\thinspace(9.24) 
of~\cite{Miransky:1994vk}.
In particular, for 
vanishing current quark mass, $m=0$, we have
\begin{align}\nonumber 
\partial^{y}_{\mu} 
\langle 0\vert T j^{\mu,a}_{5}(y) \varphi_1(x_1) \dots 
\varphi_n(x_n) \vert 0\rangle
=&\, i \delta^4 (y-x_1)
 \langle 0\vert T\frac{\partial\varphi_1(x_1)}{\partial\omega_{5\alpha}(x_1)} \cdots 
\varphi_n(x_n) \vert 0\rangle
+ \cdots\\[1ex]
&\, +
i \delta^4 (y-x_n)
\langle 0\vert T  \varphi_1(x_1) \cdots 
\frac{\partial\varphi_n(x_n)}{\partial\omega_{5\alpha}(x_n)}\vert 0\rangle \,,
\label{mastereq}
\end{align}
where
the axial-vector current,
$j^{\mu,a}_{5}(x)$, 
is given by 
\be\label{eq:j5def}
j^{\mu,a}_5(x)=\bar{\psi}(x)\gamma_5 \gamma^\mu t^a{\psi}(x)\,;
\ee
where $t^a$ are the generators of flavour $SU(N_f)$ algebra, \eg the Gell-Mann matrices for $N_f=3$; the $\varphi_i$ represent field operators of arbitrary kind, $\partial^{y}_{\mu}$ is the derivative with respect to the spacetime variable $y$, the $T$ denotes the standard time-ordering operator, and $\omega_{5\alpha}(x)$ is the angle of a local chiral transformation.

We now specialize 
\1eq{mastereq}
to the case $n=2$, where  
$\varphi_1(x_1) = \psi(x_1)$ 
and \mbox{$\varphi_1(x_2) = \bar\psi(x_2)$}.
To that end, 
we consider a local chiral transformation of the quark fields $\psi(x)$ and $\bar{\psi}(x)$ \cite{Fujikawa:1980eg}
\begin{align}
\delta \psi(x) = i 
\omega_{5\alpha}(x) \gamma_5 
t^{a} \psi(x)\,,
\qquad\qquad 
\delta {\bar\psi}(x) = i 
\omega_{5\alpha}(x)\bar\psi(x) \gamma_5 t^{a} \,,
\end{align}
and therefore, 
\begin{align}
\frac{\partial\psi(x)}{\partial\omega_{5\alpha}(x)} = 
i \gamma_5 
t^{a} \psi(x)\,,
\qquad \qquad
\frac{\partial {\bar\psi}(x)}{\partial\omega_{5\alpha}(x)} =
i {\bar\psi}(x) \gamma_5 t^{a} \,.
\label{deriv}
\end{align}
Thus, 
using \1eq{deriv},
and suppressing isospin indices, 
we have 
\begin{align}\nonumber 
\partial^{y}_{\mu} 
\langle 0\vert T j_{5}^\mu(y) 
\psi(x_1) {\bar\psi}(x_2)
\vert 0\rangle
=&\, - \delta^4 (y-x_1)
 \gamma_5\langle 0\vert T \psi(x_1) {\bar\psi}(x_2) \vert 0\rangle
 \\[1ex]
&- 
\delta^4 (y-x_2)
\langle 0\vert T  
\psi(x_1) {\bar\psi}(x_2)
\vert 0\rangle \gamma_5\,.
\label{WI1}
\end{align}
The Fourier transform of the correlation function  \mbox{$\widetilde{\Gamma}_{\!5}^\mu(y,x_1,x_2):=\langle 0|Tj^{\mu}_5(y)\psi(x_1)\bar{\psi}(x_2)|0\rangle$} defines the 
connected 
axial-vector 
vertex,
$\widetilde \Gamma_{5}^\mu(P,p_2, -p_1)$,
namely 
\be
(2\pi)^4 \delta^4(P+p_2-p_1)
\widetilde \Gamma_{5}^\mu(P,p_2, -p_1)
=
\int_{-\infty}^{+\infty} \!\!\!\!d^4 y \,d^4x_1 \, d^4 x_2 \,
e^{i(Py -p_1 x_1 + p_2 x_2)}
\widetilde{\Gamma}_{\!5}^\mu(y,x_1,x_2)\,,
\label{GammaFT}
\ee
while the Fourier transform for the 
quark propagator is given by 
\be\label{eq:quarkpropdef}
(2\pi)^4 \delta^4(p+q) iS(p) =
\int_{-\infty}^{+\infty} 
\!\!\!\!d^4x_1 d^4x_2 e^{i(p x_1 + q x_2)}
\langle 0\vert T \psi(x_1) {\bar\psi}(x_2) \vert 0\rangle\,.
\ee
We next 
integrate both sides of \1eq{WI1} by 
$\int d^4 y \,d^4x_1 \, d^4 x_2 \,
e^{i(Py -p_1 x_1 + p_2 x_2)}$. Then, on the l.h.s. 
we carry out the standard 
integration by parts, which brings in a $-iP_{\mu}$,
while on the r.h.s. we simply recover two Fourier transforms of the quark propagator. 
Thus, we get  
\be
P_{\mu}\widetilde \Gamma_{5}^\mu(P,p_2, -p_1)
= \gamma_5 S(p_1)+\gamma_5 S(p_2)
\,.
\label{wiprel}
\ee
The amputated
axial-vector vertex 
$\ga^\mu(P,p_2,-p_1)$
is defined as
\be
\widetilde \Gamma_{5}^\mu(P,p_2, -p_1)
=iS(p_1) \ga^\mu(P,p_2,-p_1) iS(p_2)\,,
\ee
and, therefore, from \1eq{wiprel} 
we get 
\be
- P^{\mu}\Gamma_{5}^\mu(P,p_2,-p_1)
= S^{-1}(p_1) \gamma_5 + \gamma_5 S^{-1}(p_2) \,,
\label{wicen}
\ee
namely the 
\wtig in \1eq{eq:WTI5}.

%%%%%%%%%%%%%%%%%%%%%%%%%%%%%
\subsection{The \texorpdfstring{$[{\rm WTI}]_{G_5}$}{wtiG5}}
\label{Appb}

In the case of the gluon-axial-vector vertex $G_5^{\mu\nu}(P,q,p_2,-q_1)$, where $q_1=p_1+q$, we consider again the master formula
of \1eq{mastereq} for $n=3$,
choosing 
$\varphi_1(x_1) = \psi(x_1)$, 
$\varphi_1(x_2) = \bar\psi(x_2)$,
and 
$\varphi_1(x_3) = A_b^\nu(x_3)$,
where $A_b^\nu(x_3)$ is a gluon field. Since 
the gluon field is invariant under the chiral transformation, we have 
$\partial A_b^\nu(x_3)(x)/\partial\omega_{5\alpha}(x) =0$. Thus,  \1eq{mastereq} yields (we suppress color indices) 
\bea
\partial^{y}_{\mu} 
\langle 0\vert T j_{5}^{\mu}(y) 
\psi(x_1) \bar\psi(x_2) A^{\nu}(x_3)
\vert 0\rangle
&=& - \delta^4 (y-x_1)
 \gamma_5\langle 0\vert T \psi(x_1) {\bar\psi}(x_2) A^{\nu}(x_3)\vert 0\rangle
\nonumber\\
&&- 
\delta^4 (y-x_2)
\langle 0\vert T  
\psi(x_1) {\bar\psi}(x_2) A^{\nu}(x_3)
\vert 0\rangle \gamma_5\,.
\label{WI2}
\eea
The Fourier transform of the connected
$\mathbb{\widetilde G}_5^{\mu\nu}(P,q,p_2,-q_1)$
is given by 
\bea
(2\pi)^4 \delta^4(P+q+p_2 -q_1)
\mathbb{\widetilde G}_5^{\mu\nu}(P,q,p_2,-q_1)
 & = & 
\displaystyle \int_{-\infty}^{+\infty} \!\!\!\!d^4 y \,d^4x_1 \, d^4 x_2 \, d^4 x_3
e^{i(Py -q_1 x_1 + p_2 x_2 + q x_3)}\nonumber\\
\nonumber\\
& \times& \displaystyle \langle 0\vert T j_{5}^{\mu}(y){\psi}(x_1)\bar{\psi}(x_2) A^{\nu}(x_3)\vert 0\rangle\,,
\eea
whereas the Fourier transform of the 
connected quark-gluon vertex 
${\widetilde\Gamma}^{\nu}(q, r, -p)$
is given by 
\be
(2\pi)^4 \delta^4(q+r-p) 
{\widetilde\Gamma}^{\nu}(q,r,-p) = \int_{-\infty}^{+\infty} \!\!\!\!d^4 y \,d^4x_1 \, d^4 x_2 \,
e^{i(-p x_1 + rx_2 +q x_3)}
\langle 0\vert {\psi}(x_1)\bar{\psi}(x_2) A^{\nu} (x_3)\vert 0\rangle \,.
\ee
Using these relations, one obtains from \1eq{WI2}
\be
iP_{\mu} \mathbb{\widetilde G}_{5}^{\mu\nu}(P,q,p_2,-q_1) = \gamma_5 {\widetilde\Gamma}^{\nu}(q,p_1,-q_1)+{\widetilde\Gamma}^{\nu}(q,p_2,-q_2)\gamma_5\,.
\label{W13}
\ee
We next introduce the amputated 
$\mathbb{G}_5^{\mu\nu}$ 
and the amputated quark-gluon vertex 
${\Gamma}^{\nu}$, 
given by
\bea
\mathbb{\widetilde G}_5^{\mu\nu}(P,q,p_2,-q_1) &=& \Delta^{\nu}_{\rho}(q)
iS(q_1) \mathbb{G}_5^{\mu\rho}(P,q,p_2,-q_1) iS(p_2)\,,
\nonumber\\
{\widetilde\Gamma}^{\nu}(q,r, -p) 
&=& \Delta^{\nu}_{\rho}(q) i S(p) 
i {\Gamma}^{\rho}(q, r, -p) i S(r)\,,
\label{newver}
\eea
where $S$ is the quark propagator defined in \1eq{eq:quarkpropdef}, and $\Delta_{\mu\nu}(q)$ is the gluon propagator.

Employing these definitions into 
\1eq{W13},  and using that 
$q_{i} = q + p_{i}$ ($i=1,2$), we get 
\bea
P_{\mu} {\mathbb G}_5^{\mu\nu}(P,q,p_2,-q_1) &=& 
\Gamma^{\nu}(q, p_1,-q_1)
S(p_1)\gamma_5 S^{-1}(p_2) \nonumber\\
&+& S^{-1}(q_1) \gamma_5 S(q_2) \Gamma^{\nu}(q, p_2,-q_2)\,.
\label{qwer1}
\eea

\begin{figure}[t!]
    \centering
    \includegraphics[scale=1]{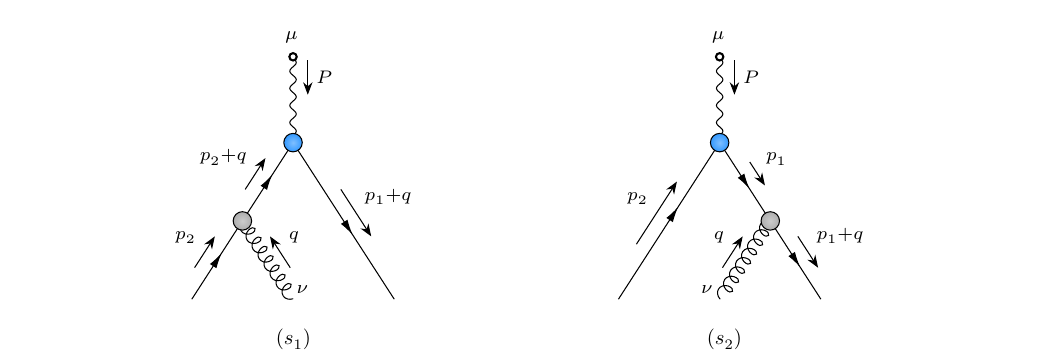}
    \caption{1PR contributions to the amputated, connected part of the gluon-axial-vector function, $\mathbb{G}_5^{\mu\nu}$. The blue (gray) circles represent fully-dressed axial-vector (quark-gluon) vertices.}
    \label{fig:1pr}
\end{figure}

We next decompose  ${\mathbb G}_5^{\mu\nu}$
into a 1PI part, denoted by 
$i G_5^{\mu\nu}$, and a 1PR part, 
denoted by ${\cal G}_5^{\mu\nu}$, 
\ie 
${\mathbb G}_5^{\mu\nu} = 
i G_5^{\mu\nu} + 
{\cal G}_5^{\mu\nu}$. The contributions composing 
${\cal G}_5^{\mu\nu}$
originate from diagrams 
$(s_1)$ and $(s_2)$ in \fig{fig:1pr},
namely 
${\cal G}_5^{\mu\nu} = 
(s_1)_5^{\mu\nu} + (s_2)_5^{\mu\nu}$, which 
are given by 
\bea
(s_1)_5^{\mu\nu} &=&  
\Gamma_5^{\mu}(P,p_2,-q_1) iS(q_2) 
i\Gamma^{\nu} (q,p_2,-q_2) \,,
\nonumber\\
(s_2)_5^{\mu\nu} &=&
i\Gamma^{\nu} (q,p_1,-q_1) iS(p_1) \Gamma_{5}^{\mu}
(P,p_2,-p_1) \,.
\label{theeses}
\eea
Then, using \1eq{wicen}, we find 
\begin{align}\nonumber
P_{\mu}(s_1)_5^{\mu\nu} =&\,
S^{-1}(q_1) \gamma_5 S(q_2) \Gamma^{\nu}(q, p_2,-q_2) + 
\gamma_5 \Gamma^{\nu}(q, p_2,-q_2)
\\[1ex]
P_{\mu}(s_2)_5^{\mu\nu} =&\,
\Gamma^{\nu}(q,p_1,-q_1)
\gamma_5 + 
\Gamma^{\nu}(q,p_1,-q_1)
S(p_1)\gamma_5 S^{-1}(p_2) \,.
\label{s1s2}
\end{align}
We note now that the terms in 
\1eq{s1s2} contain 
the r.h.s. of 
\1eq{qwer1}, namely 
\begin{align}
 P_{\mu} {\cal G}_5^{\mu\nu} 
 = 
 \Gamma^{\nu}(q,p_1,-q_1)
\gamma_5 + 
\gamma_5 \Gamma^{\nu}(q,p_2,-q_2)
+ {\rm [r.h.s.]_{\eqref{qwer1}} }
\end{align}
and therefore, from \1eq{newver}
we get 
\begin{align}
i P_{\mu} G_5^{\mu\nu} +
\Gamma^{\nu}(q,p_1,-q_1)
\gamma_5 + 
\gamma_5 \Gamma^{\nu}(q,p_2,-q_2) = 0 
\end{align}
which is precisely the 
\wtiG of \1eq{eq:G5WTI}.

\section{Flavour singlet and non-singlet currents}\label{app:singletnonsinglet}

Throughout our construction we have considered the axial-vector current, which behaves as a non-singlet under flavour rotations  $U\in SU(N_f)$. Specifically, the quark fields transform as
\begin{align}
    \psi(x)&\to U\psi(x)\,,& \bar{\psi}(x) &\to\bar{\psi}(x) U^\dag\,,
\end{align}
and therefore, the axial-vector current defined in \1eq{eq:j5def} transforms as
\be
j_5^{\mu,a}(x)\to \bar{\psi}(x)U^\dag \gamma_5\gamma^\mu t^a U\psi(x)\,.
\ee
Since
\be
U^\dag t^a U = R^{ab}(U)t^b
\ee
where $R^{ab}(U)$ defines the adjoint representation of the flavour group, 
we have that 
\be
j_5^{\mu,a}(x)\to R^{ab}(U)j_5^{\mu,b}(x)\,,
\ee
which is a flavour non-singlet quantity. In contrast,
one may show following the same procedure 
that the current $j_5^{\mu,s}(x)=\bar{\psi}(x)\gamma_5\gamma^\mu\psi(x)$ 
is a flavour singlet.

\begin{figure}
    \centering
    \includegraphics[scale=1]{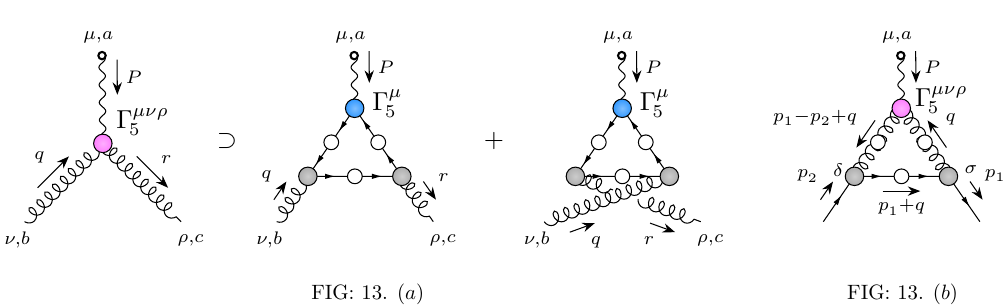}
    \caption{\textit{Left panel}: Gluon-gluon-axial-vector vertex; \textit{Right panel}: Gluon-gluon-axial-vector vertex contribution to the singlet axial-vector vertex SDE.}
    \label{fig:singlet}
\end{figure}

Since gluons do not 
couple to flavour non-singlets, the 
diagram shown on \fig{fig:singlet}. ($b$) is absent from the 
SDE of the non-singlet axial-vector vertex
$t^a\Gamma_5^{\mu\nu}$, see \fig{fig:SDEPicture}.  
Instead, it must be added to the SDE of a 
singlet axial-vector vertex, see \eg \cite{Eichmann:2016yit}.

Note that the non-singlet current is anomaly-free, see \eg \cite{Miransky:1994vk}, and thus is conserved in the chiral limit; in contrast, the singlet 
current is not conserved in the same limit,
displaying the Adler-Bell-Jackiw anomaly \cite{Schwinger:1951nm, Schwinger:1962tp, Adler:1969gk, Bell:1969ts, Bardeen:1969md}. Specifically, 
the vertex ${\g}_{\!5}^{\mu\nu\rho}(P,q,-r)$, \fig{fig:singlet}. ($a$),
which may be defined from the 
Fourier transform of the correlation function $\langle 0\vert Tj^{\mu}_5(y)A^\nu(x_1)A^\rho(x_2)\vert 0\rangle$, in complete analogy to 
\1eq{GammaFT}, is transverse up the anomaly, that is 
\be
P_{\mu} {\g}_{\!5}^{\mu\nu\rho}(P,q,-r) = 0 + \frac{ig^2}{16\pi^2}A^{\nu\rho}(P,q,-r) \,,
\ee
where the quantity $A^{\nu\rho}(P,q,-r)$ is defined from the Fourier transform of the anomalous term 
\mbox{$\langle 0\vert T\epsilon^{\mu\sigma\kappa\delta}\textrm{Tr}[F_{\kappa\delta}(x)F_{\mu\sigma}(x)]A^\nu(x_1)A^\rho(x_2)\vert 0\rangle$}, see, \eg  \cite{Miransky:1994vk}.

\color{black}

\bibliography{bibliography}

\end{document}